# Guidelines for the Development of Immersive Virtual Reality Software for Cognitive Neuroscience and Neuropsychology: The Development of Virtual Reality Everyday Assessment Lab (VR-EAL), A Neuropsychological Test Battery in Immersive Virtual Reality.

**Panagiotis Kourtesis[1,2,3,4], Danai Korre[5], Simona Collina[3,4], Leonidas A.A. Doumas[2], and Sarah E. MacPherson[1,2]**

[1] Human Cognitive Neuroscience, Department of Psychology, University of Edinburgh, Edinburgh, United Kingdom

[2] Department of Psychology, University of Edinburgh, Edinburgh, United Kingdom

[3] Lab of Experimental Psychology, Suor Orsola Benincasa University of Naples, Naples, Italy

[4] Interdepartmental Centre for Planning and Research "Scienza Nuova", Suor Orsola Benincasa University of Naples, Naples, Italy

[5] Centre for Intelligent Systems and their Applications, School of Informatics, University of Edinburgh, Edinburgh, United Kingdom

**\* Correspondence:**
Panagiotis Kourtesis
pkourtes@exseed.ed.ac.uk

**Abstract**

Virtual reality (VR) head-mounted displays (HMD) appear to be effective research tools, which may address the problem of ecological validity in neuropsychological testing. However, their widespread implementation is hindered by VR induced symptoms and effects (VRISE) and the lack of skills in VR software development. This study offers guidelines for the development of VR software in cognitive neuroscience and neuropsychology, by describing and discussing the stages of the development of Virtual Reality Everyday Assessment Lab (VR-EAL), the first neuropsychological battery in immersive VR. Techniques for evaluating cognitive functions within a realistic storyline are discussed. The utility of various assets in Unity, software development kits, and other software are described so that cognitive scientists can overcome challenges pertinent to VRISE and the quality of the VR software. In addition, this pilot study attempts to evaluate VR-EAL in accordance with the necessary criteria for VR software for research purposes. The VR neuroscience questionnaire (VRNQ; Kourtesis et al., 2019b) was implemented to appraise the quality of the three versions of VR-EAL in terms of user experience, game mechanics, in-game assistance, and VRISE. Twenty-five participants aged between 20 and 45 years with 12-16 years of full-time education evaluated various versions of VR-EAL. The final version of VR-EAL achieved high scores in every sub-score of the VRNQ and exceeded its parsimonious cut-offs. It also appeared to have better in-game assistance and game mechanics, while its improved graphics substantially increased the quality of the user experience and almost eradicated VRISE. The results substantially support the feasibility of the development of effective VR research and clinical software without the presence of VRISE during a 60-minute VR session.

## 1    Introduction

In cognitive neuroscience and neuropsychology, the collection of cognitive and behavioral data is predominantly achieved by implementing psychometric tools (i.e., cognitive screening and testing). The psychometric tools are principally limited to paper-and-pencil and computerized (i.e., 2D and 3D applications) forms. Psychometric tools in clinics and/or laboratories display several limitations and discrepancies between the observed performance in the laboratory/clinic and the actual performance of individuals in everyday life (Rizzo *et al.*, 2004; Bohil *et al.*, 2011; Parsons, 2015). The functional and predictive association between an individual's performance on a set of neuropsychological tests and the individual's performance in various everyday life settings is called ecological validity. Ecological validity is considered an important issue that cannot be resolved by the currently available assessment tools (Rizzo *et al.*, 2004; Bohil *et al.*, 2011; Parsons, 2015).

Ecological validity is especially important in the assessment of certain cognitive functions, which are crucial for performance in everyday life (Chaytor & Schmitter-Edgecombe, 2003). In particular, executive functioning (e.g., multitasking, planning ability and mental flexibility) has been found to predict occupational and academic success (Burgess *et al.*, 1998). Similarly, the ecologically valid measurement of memory (e.g., episodic memory) and attentional processes (e.g., selective, divided, and sustained attention) have been seen as predictors of overall performance in everyday life (Higginson *et al.*, 2000). Lastly, prospective memory (i.e., the ability to remember to carry out intended actions at the correct point in the future; McDaniel & Einstein, 2007) plays an important role in everyday life and the assessment of prospective memory abilities requires ecologically valid tasks (Phillips *et al.*, 2012).

Current ecologically valid tests are not thought to encompass the complexity of real-life situations (Rizzo *et al.*, 2004; Bohil *et al.*, 2011; Parsons, 2015). Assessments which take place in real-world settings (e.g., performing errands in a shopping center) are time consuming and expensive to set up, lack experimental control over the external situation (e.g., Elkind *et al.*, 2001), cannot be standardized for use in other labs, and are not feasible for certain populations (e.g., individuals with psychiatric conditions or motor difficulties) (Rizzo *et al.*, 2004; Parsons, 2015). The traditional approaches in cognitive sciences encompass the employment of static and simple stimuli, which lack ecological validity. Instead, immersive virtual reality (VR) technology enables cognitive scientists to accumulate advanced cognitive and behavioral data through the employment of dynamic stimuli and interactions with a high degree of control within an ecologically valid environment (Rizzo *et al.*, 2004; Bohil *et al.*, 2011; Parsons, 2015). Furthermore, VR can be combined with non-invasive imaging techniques (Makeig *et al.*, 2009; Bohil *et al.*, 2011; Parsons, 2015), wearable mobile brain/body imaging (Makeig *et al.*, 2009), and can be used for rehabilitation and treatment purposes (Rizzo *et al.*, 2004; Bohil *et al.*, 2011; Parsons, 2015).

VR has great potential as an effective telemedicine tool that may resolve the current methodological problem of ecological validity (Rizzo *et al.*, 2004; Bohil *et al.*, 2011; Parsons, 2015; Parsons *et al.*, 2018). However, the appropriateness of VR, especially for head-mounted display (HMD) systems, is still controversial (Bohil *et al.*, 2011; de Franca & Soares, 2017; Palmisano *et al.*, 2017). The principal concern is the adverse symptomatology (i.e., nausea, dizziness, disorientation, fatigue, and instability) which stems from the implementation of VR systems (Bohil *et al.*, 2011; de Franca & Soares, 2017; Palmisano *et al.*, 2017). These adverse VR induced symptoms and effects (VRISE) endanger the health and safety of the users (Parsons *et al.*, 2018), decrease reaction times and overall cognitive performance (Nalivaiko *et al.*, 2015), while increasing body temperature and heart rates (Nalivaiko *et al.*, 2015), cerebral blood flow and oxyhemoglobin concentration (Gavgani *et al.*, 2018), brain activity (Arafat *et al.*, 2018), and the connectivity between brain regions (Toschi *et al.*, 2017). Hence, VRISE may compromise the reliability of cognitive, physiological, and neuroimaging data (Kourtesis *et al.*, 2019a).

However, VRISE predominantly stem from hardware and software inadequacies, which more contemporary commercial VR hardware and software do not share (Kourtesis *et al.*, 2019a; Kourtesis *et al.*, 2019b). The employment of modern VR HMDs analogous to or more cutting-edge than the HTC Vive and/or Oculus Rift, in combination with ergonomic VR software, appear to significantly mitigate the presence of VRISE (Kourtesis *et al.*, 2019a; Kourtesis *et al.*, 2019b). However, the selection of suitable VR hardware and/or software demands acceptable technological competence (Kourtesis *et al.*, 2019a). Minimum hardware and software features have been suggested to appraise the suitability of VR hardware and software (Kourtesis *et al.*, 2019a). The technical specifications of the computer and VR HMD are adequate to assess their quality (Kourtesis *et al.*, 2019a), while the virtual reality neuroscience questionnaire (VRNQ) facilitates the quantitative evaluation of software attributes and the intensity of VRISE (Kourtesis *et al.*, 2019b).

Another limitation is that the implementation of VR technology may necessitate high financial costs, which hinders its widespread adoption by cognitive scientists. In the 90s, the cost of a VR lab with basic features cost between $20,000 and $50,000, where nowadays the cost has decreased considerably (Slater, 2018). At present, the cost of a VR lab with basic features (e.g., a HMD, external hardware, and laptop) is between $2,000 and $2,500. However, the development of VR software is predominantly dependent on third parties (e.g., freelancers or companies) with programming and software development skills (Slater, 2018). A solution that will promote the adoption of immersive VR as a research and clinical tool might be the in-house development of VR research/clinical software by computer science literate cognitive scientists or research software engineers.

The current study endeavors to offer guidelines on the development of VR software by presenting the development of the Virtual Reality Everyday Assessment Lab (VR-EAL). Since the assessment of prospective memory, episodic memory, executive functions, and attention are likely to benefit from ecologically valid approaches to assessment, VR-EAL attempts to be one of the first neuropsychological batteries to apply immersive VR to assess these cognitive functions. However, the ecologically valid assessment of these cognitive functions demands the development of a realistic scenario with several scenes and complex interactions while avoiding intense VRISE factors.

The VR-EAL development process is presented systematically, aligned with the steps that cognitive scientists should follow to achieve their aim of designing VR studies. Firstly, the preparation stages are described and discussed. Secondly, the structure of the application (e.g., order of the scenes) is presented and discussed in terms of offering comprehensive tutorials, delivering a realistic storyline, and incorporating a scoring system. Thirdly, a pilot study is conducted to evaluate the suitability of the different versions of VR-EAL (i.e., alpha, beta, final) for implementation in terms of user experience, game mechanics, in-game assistance, and VRISE.

## 2    Development of VR-EAL

### 2.1    Rationale and Preparation

Prospective memory encompasses the ability to remember to initiate an action in the future (Anderson *et al.*, 2017). The prospective memory action may be related to a specific event (e.g., when you see this person, give him a particular object) or time (e.g., at 5 pm perform a particular task). Attentional control processes, executive functioning, the difficulty of the filler/distractor tasks, the length of the delay between encoding the intention to perform a task and the presentation of the stimulus-cue, as well as the length of the ongoing task, all affect prospective memory ability (Anderson *et al.*, 2017). Therefore, the VR-EAL scenarios need to incorporate both types of prospective memory actions and consider the length and difficulty of the distractor tasks and delays, as well as attentional and executive functioning.

The main theoretical frameworks of prospective memory are the preparatory attentional and memory (PAM) and the multiprocess (MP) theories (Anderson *et al.*, 2017). The PAM theory suggests that performing prospective memory tasks efficiently requires a constant top-down monitoring for environmental and internal cues in order to recall the intended action and perform it (Smith, 2003; Smith *et al.*, 2007). For example, an individual wants to buy a pint of milk after work. On her way home, she is vigilant (i.e., monitoring) about recognizing prompts (e.g., the sign of a supermarket) that will remind her of her intention to buy a pint of milk. In addition to PAM's top-down monitoring, MP theory suggests that bottom-up spontaneous retrieval also enables effective performance on prospective memory tasks (McDaniel & Einstein, 2000; McDaniel & Einstein, 2007). Going back to the previous example, when the individual is not being vigilant (i.e., passive), she sees an advert pertaining to dairy products, which triggers the retrieval of her intention to buy a pint of milk. VR-EAL is required to incorporate both predominant retrieval strategies in line with these main theoretical frameworks of prospective memory (i.e., PAM and MP). This may be achieved by including scenes where the user should be vigilant (i.e., PAM) so they recognize a stimulus associated with the prospective memory task (e.g., notice a medicine on the kitchen's table in order to take it after having breakfast), as well as scenes where the user passively (i.e., MP) will attend to an obvious stimulus related to the prospective task (e.g., while being in front of the library, the user needs to remember to return a book).

Notably, the ecologically valid assessment of executive (i.e., planning and multitasking), attentional (i.e., selective visual, visuospatial, and auditory attention), and episodic memory processes is an equally important aim of VR-EAL. The relevant literature postulates that the everyday functioning of humans is dependent on cognitive abilities, such as attention, episodic memory, prospective memory, and executive functions (Chaytor & Schmitter-Edgecombe, 2003; Haines et al., 2019; Higginson et al., 2000; Mlinac & Feng, 2016; Phillips et al., 2012; Rosenberg, 2015). However, the assessment of these cognitive functions requires an ecologically valid approach to indicate the quality of the everyday functioning of the individual in the real world (Chaytor & Schmitter-Edgecombe, 2003; Haines et al., 2019; Higginson et al., 2000; Mlinac & Feng, 2016; Phillips et al., 2012; Rosenberg, 2015). However, the assessment (i.e., tasks) of these cognitive functions in VR-EAL will also serve as distractor tasks for the prospective memory components of the paradigm. Hence, the VR-EAL distractor tasks are vital to the prospective memory tasks, but at the same time, they are adequately challenging within a continuous storyline (see Table 1).

Furthermore, ecologically valid tasks performed in VR environments demand various game mechanics and controls to facilitate ergonomic and naturalistic interactions, and these need to be learnt by users. The scenario should include tutorials that allow users to spend adequate time learning how to navigate, use and grab items, and how the VE reacts to his/her actions (Gromala *et al.*, 2016; Jerald *et al.*, 2017; Brade *et al.*, 2018) (see Table 1). Additionally, the scenario should consider the in-game instructions and prompts offered to users such as directional arrows, non-player characters (NPC), signs, labels, ambient sounds, audio, and videos that aid performance (Gromala *et al.*, 2016; Jerald *et al.*, 2017; Brade *et al.*, 2018). Importantly, this user-centered approach appears to particularly favor non-gamers in terms of performing better and enjoying the VR experience (Zaidi *et al.*, 2018). Thus, the development of VR-EAL should be aligned with these aforementioned suggestions.

The first step of the development process was to select the target platform. In VR's case, this is the VR HMD, which allows various interactions to take place within a virtual environment (VE) during the neuropsychological assessment. In our previous work (Kourtesis *et al.*, 2019a), we have highlighted a number of suggested minimum hardware and software features which appraise the suitability of VR hardware and software. Firstly, interactions with the VE should be ergonomic in order to elude or alleviate the presence of VRISE. Also, the utilization of 6 degrees of freedom (DoF) wands (i.e., controllers) facilitates ergonomic interactions and provides highly accurate motion tracking. Lastly, the two types of HMD that exceed the minimum standards and support 6DoF controllers are the HTC Vive and Oculus Rift; hence, the target HMD should have hardware characteristics equal to or greater than these high-end HMDs (Kourtesis *et al.*, 2019a). VR-EAL is developed to be compatible with HTC Vive, HTC Vive Pro, Oculus Rift, and Oculus Rift-S.

The second step was to select which game engine (GE) should be used to develop the VR software. For the development of VR-EAL, the feasibility of acquiring the required programming and software development skills was an important criterion for the selection of the GE because the developer of VR-EAL (i.e., the corresponding author) is a cognitive scientist who did not have any background in programming or software development. The two main GEs are Unity and Unreal. Unity requires C# programming skills, while Unreal requires C++ programming skills. Learners of C#, either experienced or inexperienced programmers, appear to experience a greater learning curve than learners of C++ (Chandra, 2012). While Unity and Unreal are of equal quality (Dickson *et al.*, 2017), Unity as a GE has been found to be more user-friendly, and easier to learn compared to Unreal (Dickson *et al.*, 2017). Also, Unity has an extensive online community and online resources (e.g., 3D models, software development kits; SDK), and documentation (Dickson *et al.*, 2017). For these reasons, Unity was preferred for the development of VR-EAL. However, either Unreal or Unity would have been a sensible choice since both GEs offer high quality tools and features for software development (Disckson *et al.*, 2017).

The final step was the acquisition of skills and knowledge. A cognitive scientist with a background either in computer or psychological sciences should have knowledge of the cognitive functions to be studied, as well as, intermediate programming and software development skills pertinent to the GE. The acquisition of these skills enables the cognitive scientist to design the VR software in agreement with the capabilities of the GE and the research aims. In VR-EAL's case, its developer meticulously studied the established ecologically valid paper-and-pencil tests such as the Test of Everyday Attention (TEA; Robertson *et al.*, 1996), the Rivermead Behavioral Memory Test – III (RBMT-III; *Wilson et al.*, 2008), the Behavioral Assessment of the Dysexecutive Syndrome (BADS; Wilson *et al.*, 1996), and the Cambridge Prospective Memory Test (CAMPROMPT; Wilson *et al.*, 2005). In addition, other research and clinical software were considered. For example, the Virtual Reality Shopping Task (Canty *et al.*, 2014), Virtual Reality Supermarket (Grewe *et al.*, 2014), Virtual Multiple Errands Test (Rand *et al.*, 2009), the Invisible Maze Task (Gehrke *et al.*, 2018), and the Jansari Assessment of Executive Function (Jansari *et al.*, 2014) are non-immersive VR software which assess cognitive functions such as executive functions, attentional processes, spatial cognition, and prospective memory.

Finally, the developer of VR-EAL attained intermediate programming skills in C# and software development skills in Unity. This was predominantly achieved by attending online specializations and tutorials on websites such as Coursera, Udemy, CodeAcademy, SoloLearn, and EdX. Also, a developer may consider established textbooks such as the "The VR book: Human-centered design for virtual reality" (Jerald, 2015), "3D user interfaces: theory and practice" (LaViola et al, 2017), and "Understanding virtual reality: Interface, application, and design" (Sherman & Craig, 2018). To sum it up, the acquisition of these skills enabled progression to the next stage of the development of VR-EAL, which is the writing of the scenarios/scripts.

Table 1. VR-EAL Scenario

| Order | Type | Description |
|---|---|---|
| Scene 1 | *Tutorial* | Basic interactions and navigation |
| Scene 2 | *Tutorial* | Interactive boards (recognition and planning) |
| Scene 3 | Storyline | List of prospective memory tasks, shopping list (immediate recognition), and itinerary (planning) |
| Scene 4 | *Tutorial* | List of mechanics for the prospective memory tasks, prompts, and notes |
| Scene 5 | *Tutorial* | Cooking |
| Scene 6 | Storyline | Prepare breakfast (multi-tasking) and take medication (prospective memory, event-based, short delay) |
| Scene 7 | *Tutorial* | Tutorial: collect items |
| Scene 8 | Storyline | Collect items from the living-room (selective visuospatial attention) and take a chocolate pie out of the oven (prospective memory, event-based, short delay) |
| Scene 9 | *Tutorial* | Interaction with 3D non-player characters |
| Scene 10 | Storyline | Call Rose (prospective memory task, time-based, short delay) |
| Scene 11 | *Tutorial* | Gaze interaction |
| Scene 12 | Storyline | Detect posters on both sides of the road (selective visual attention) |
| Scene 13 | *Tutorial* | Shopping, how to collect the items from the supermarket |
| Scene 14 | Storyline | Collect the shopping list items from the supermarket (delayed recognition) |
| Scene 15 | Storyline | Go to the bakery to collect the carrot cake (prospective memory task, time-based, medium delay) |
| Scene 16 | Storyline | False prompt before going to the library (prospective memory task, event-based, medium delay) |
| Scene 17 | Storyline | Return the red book to the library (prospective memory task, event-based, medium delay) |
| Scene 18 | *Tutorial* | Auditory interaction |
| Scene 19 | Storyline | Detect sounds from both sides of the road (selective auditory attention) |
| Scene 20 | Storyline | False prompt before going back home (prospective memory task, time-based, long delay) |
| Scene 21 | Storyline | When you return home, give the extra pair of keys to Alex (prospective memory task, event-based, long delay) |
| Scene 22 | Storyline | Put away the shopping items and take the medication (prospective memory task, time-based, long delay) |

## 2.2 Tutorials and mechanics

VR-EAL commences with two tutorial scenes. The first tutorial allows the user to learn how to navigate using teleportation, to hold and manipulate items (e.g., throwing them away), how to use items (e.g., pressing a button), as well as to familiarize themselves with the in-game assistance objects (e.g., a directional arrow or a sign; see Figure 1). The user is prompted to spend adequate time learning the basic interactions and navigation system because these game mechanics and in-game assistance methods are fundamental to most scenes in VR-EAL.

The second tutorial instructs the user how to use interactive boards (i.e., use a map or select items from a list). This tutorial is specific to the tasks that the user should perform in the subsequent storyline scene. Similarly, the remaining tutorials are specific to their subsequent scene (i.e., the actual task) in which the user is assessed. This design enables the user to perform the tasks, without providing them with an overwhelming amount of information that may confuse the user. However, the tutorial in the fourth scene is specific to the prospective memory tasks that are performed in several scenes throughout the scenario. The instructions for all prospective memory tasks (i.e., what should be performed and when) are provided during the third scene (i.e., storyline-bedroom scene), but the first prospective memory task is not performed until the sixth scene (i.e., the cooking task; see Table 1).

In scene 4, the user learns how to use a VR digital watch, use prospective memory items and notes (toggle on/off the menu), and follow prospective memory prompts. These game mechanics are essential to successfully perform the prospective memory tasks. The VR digital watch is the main tool for checking the time in relation to the time-based prospective memory tasks, while the prospective memory notes are crucial reminders for the time- and event-based prospective memory tasks. Subsequently, in scene 5, the user completes a tutorial where s/he learns how to use the oven and the stove as well as the snap-drop-zones to perform the cooking task. The snap-drop-zones are game objects, which are containers that the user may attach other game objects too. In scene 7, the user learns how to collect items using the snap-drop-zones attached to the left controller (see Figure 2).

In scene 9, the user learns how to interact with the 3D non-player characters (NPC). The user is required to talk to the NPC to initiate a conversation (i.e., detection of a sound through the mic input), and use the interactive boards to select a response, which either presents a dichotomous choice (i.e., "yes" or "no") or a list of items (see Figure 2). These interactions with the NPC are central to the assessment of prospective memory, and the user should effectively interact with the NPC in six scenes to successfully perform an equal number of time- and event-based prospective memory tasks.

In scene 11, the user learns how to use gaze interactions. There is a circular crosshair, which indicates the collision point of a ray that is emitted from the center of the user's visual field. The user is required to direct the circular crosshair over the targets and avoid the distractors (see Figure 2). The user needs to effectively perform a practice trial to proceed to the next scene. The practice trial requires the user to spot the three targets and avoid all the distractors while moving. If the user is unsuccessful, then the practice trial is re-attempted. This procedure is repeated until the user effectively completes the practice trial.

Scene 13 is a short tutorial where the user is reminded how to collect items using the snap-drop-zones attached to the left controller and remove an item from the snap-drop-zone in cases where an item is erroneously picked up. In scene 18, the user learns how to detect target sounds (i.e., a bell) and avoid distractors (i.e., a high-pitched and a low-pitched bell). The user looks straight ahead and presses the trigger button on the right controller when a target sound is heard on the right side. Likewise, the user presses the trigger button on the left controller when a target sound is heard on the left side (see Figure 2). The sounds are activated by trigger-zones, which are placed within the itinerary of the user. This tutorial is conducted in a similar way to the scene 11 tutorial (i.e., gaze interaction). The user, while being on the move, needs to detect three target sounds and avoid the distractors to proceed to the next scene.

The time spent on each tutorial is recorded to provide the learning time for the various interaction systems (i.e., game mechanics). However, in the scene 11 and 18 tutorials, the practice trial times are also recorded. The collected data (i.e., time spent on tutorials and the attempts to complete the practice trials) for each tutorial are added to a text file that contains the user's data (i.e., performance scores on every task).

Figure 1. VR-EAL Tutorials: Scenes 1 - 5

**Scene 1**

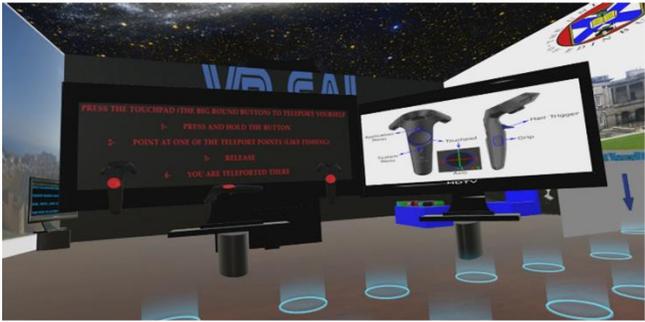

**Scene 1**

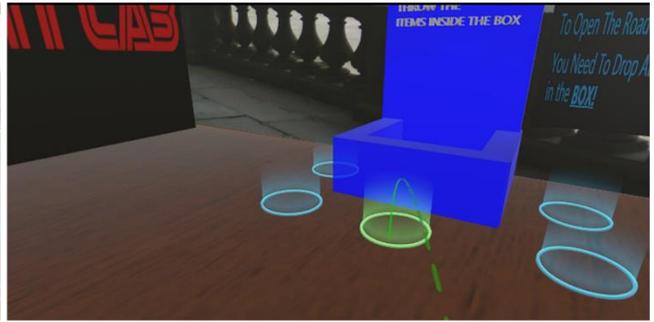

**Scene 1**

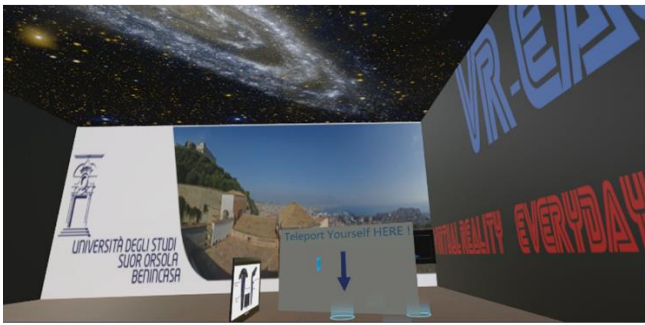

**Scene 2**

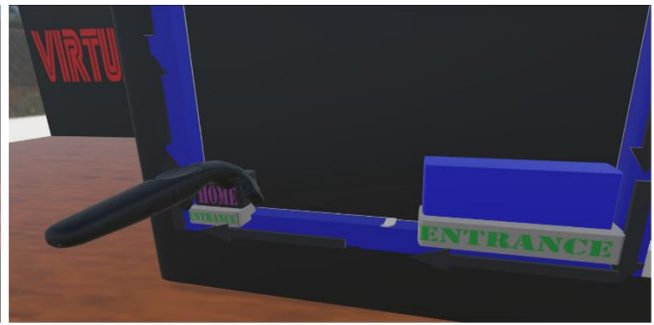

**Scene 2**

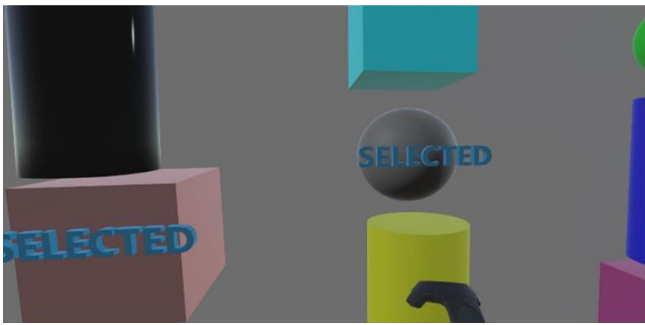

**Scene 4**

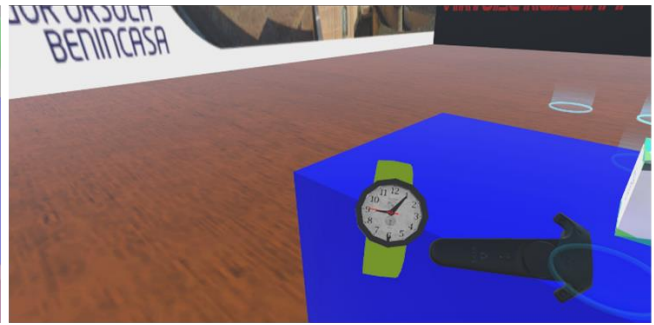

**Scene 4**

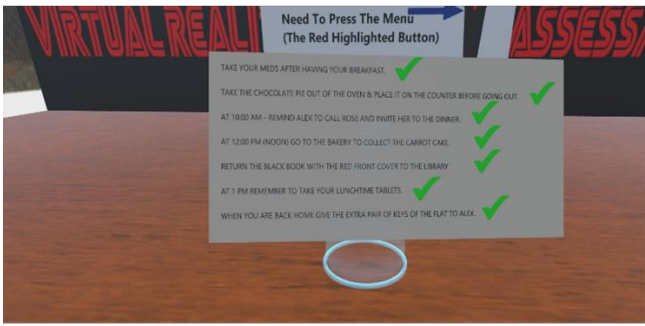

**Scene 5**

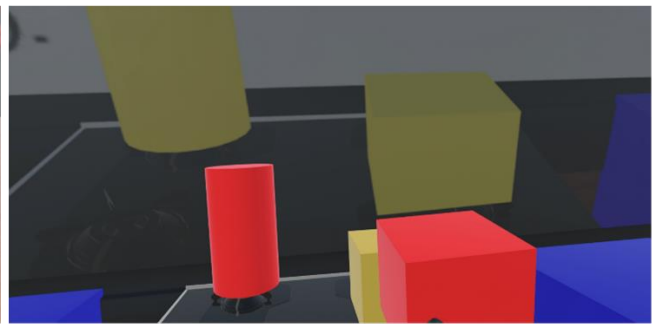

Figure 2. VR-EAL Tutorials: Scenes 7 - 18

**Scene 7**

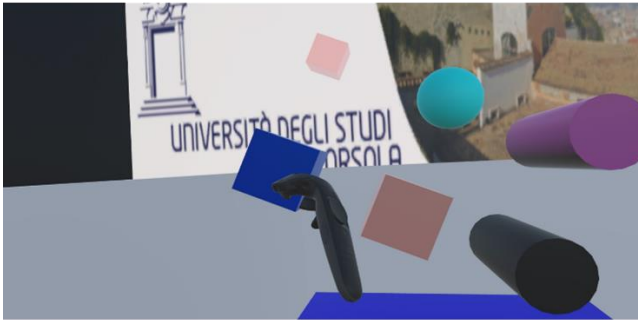

**Scene 9**

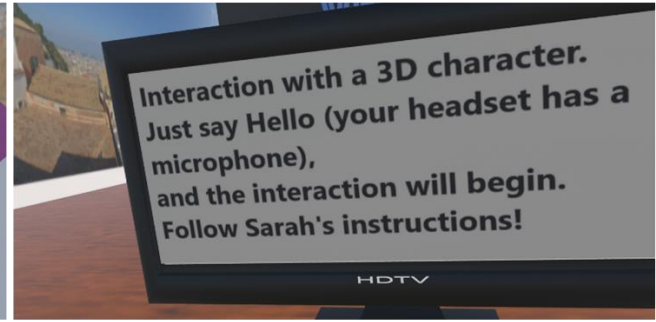

**Scene 9**

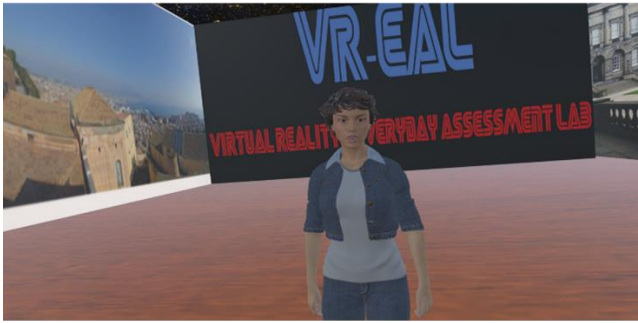

**Scene 9**

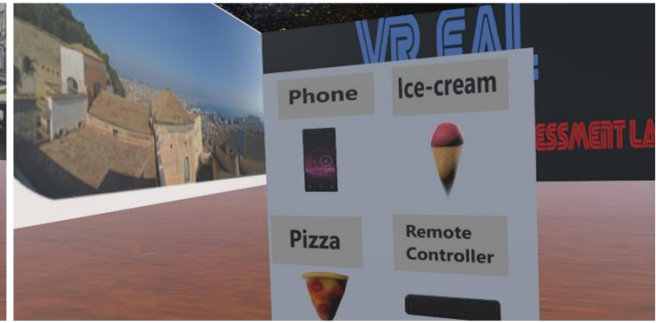

**Scene 11**

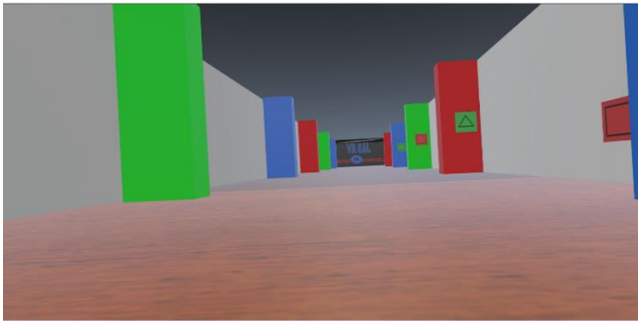

**Scene 11**

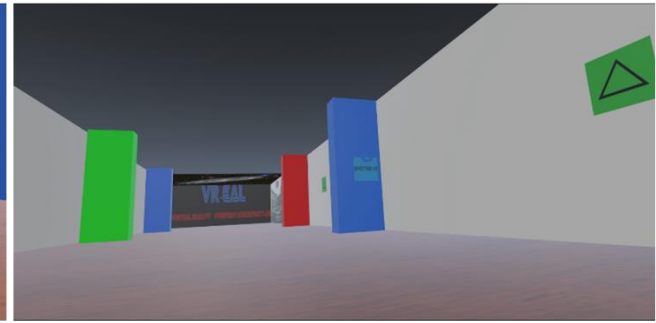

**Scene 18**

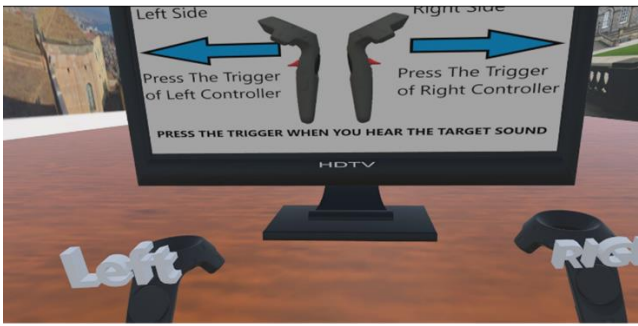

**Scene 18**

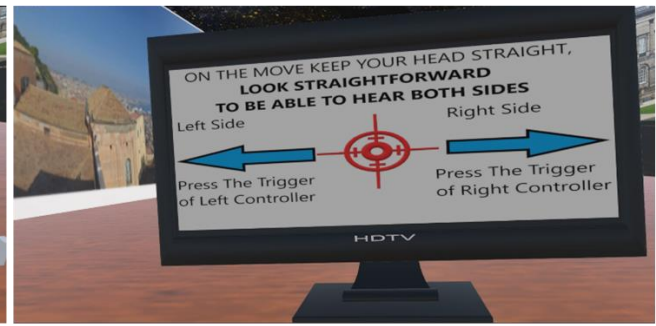

## 2.3   Storyline and Scoring

The required times to complete scenes and tasks are recorded. However, the task times are measured independently from the total scene times. Additionally, in the scenes where the user should perform prospective memory tasks, the number of times and the duration that the prospective memory notes appeared are also measured. These variables indicate how many times the user relies on the prospective memory notes, and how long they read them for.

### 2.3.1 At home

### 2.3.1.1 Bedroom: Immediate recognition and planning

The storyline commences in a bedroom (i.e., scene 3; see Table 1), where the user receives an incoming call from his/her close friend, Sarah, asking the user to carry out some errands for her (e.g., buy some shopping from the supermarket, collect a carrot cake from the bakery, return a library book). All the errands are prospective memory tasks except the shopping task. In this scene, the user should perform three different tasks. The first task is the prospective memory notes (i.e., PM-Notes) task, where the user responds affirmatively or negatively to three prompts asking the user to write down the errands (i.e., PM-tasks). The response of the user indicates his/her intention to use external tools (i.e., notes) as reminders.

The second task is the immediate recognition task where the user should choose the ten target items (i.e., create the shopping list) from an extensive array of items (see Figure 3), which also contains five qualitative distractors (e.g., semi-skimmed milk versus skimmed milk), five quantitative distractors (e.g., 1 kg potatoes versus 2 kg potatoes), and ten false items (e.g., bread, bananas etc.). The user gains 2 points for each correctly chosen item, 1 point for choosing a qualitative or quantitative distractor, and 0 points for the false items. The maximum possible score is 20 points.

The third task is the planning task. The user should draw a route on a map to visit three destinations (i.e., the supermarket, bakery and library) before returning home. The road system comprises 23 street units (see Figure 3). When the user selects a unit, 1 point is awarded. The ideal route to visit all three destinations is 15 units; hence, any extra or missing units are subtracted from the total possible score of 15. For example, if the user draws 18 units, then the distance from the ideal route is calculated as 3 (i.e., 18 - 15 = 3). Three is then subtracted from the ideal score of 15, resulting in a score of 12. If the user draws 12 units, the distance from the ideal route is also 3 and again 3 is subtracted from 15, resulting in a score of 12. The planning task score is also modified by the planning task completion time (e.g., a completion time 2 standard deviations below the average time of the normative sample is awarded 2 points while 2 standard deviations above the average time is subtracted 2 points).

Figure 3. VR-EAL Storyline: Scenes 3 - 12

**Scene 3**                                 **Scene 3**

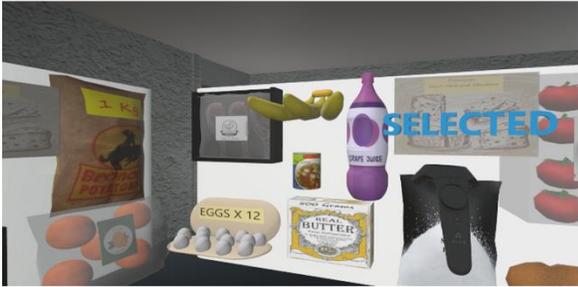 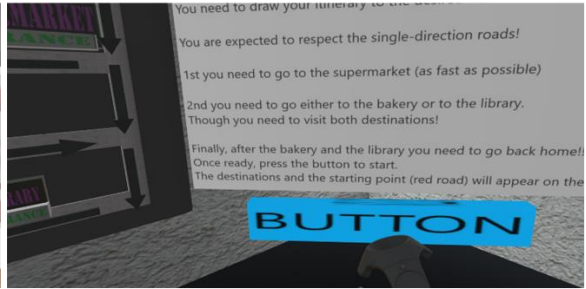

**Scene 6**                                 **Scene 6**

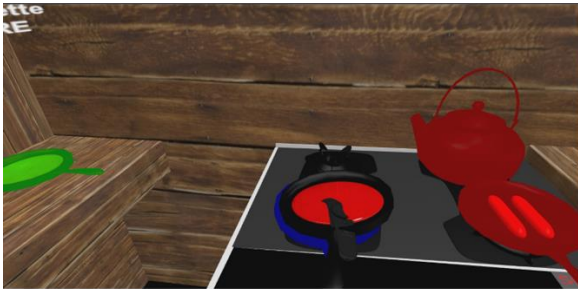 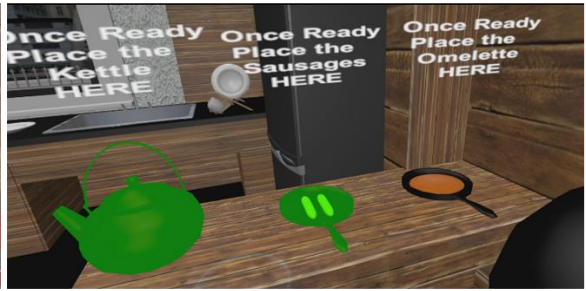

**Scene 6**                                 **Scene 8**

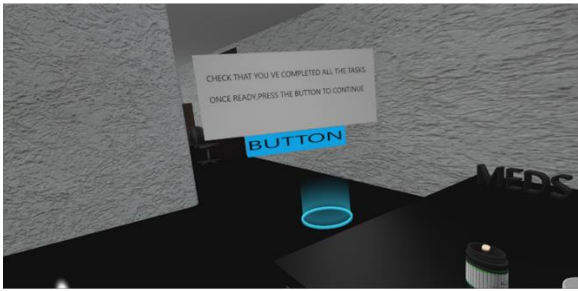 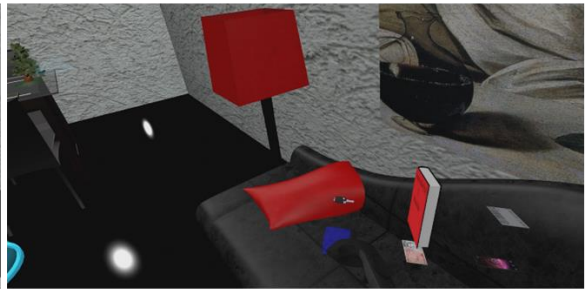

**Scene 8**                                 **Scene 10**

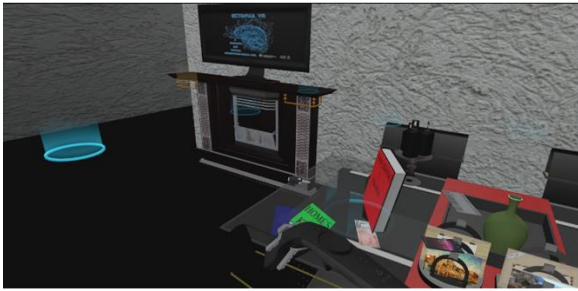 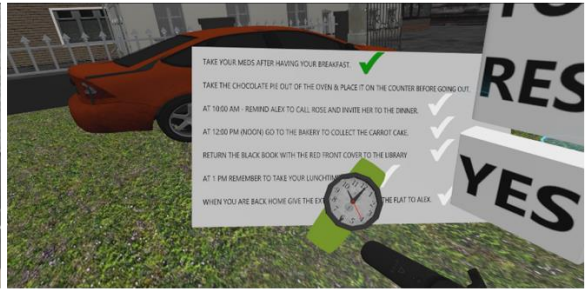

**Scene 12**                               **Scene 12**

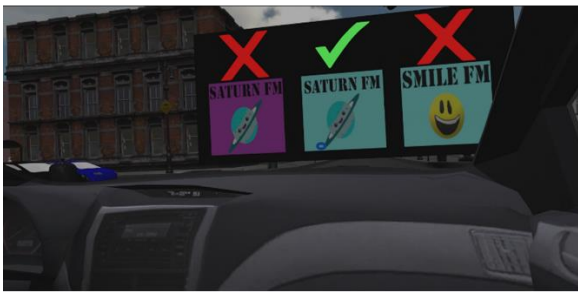 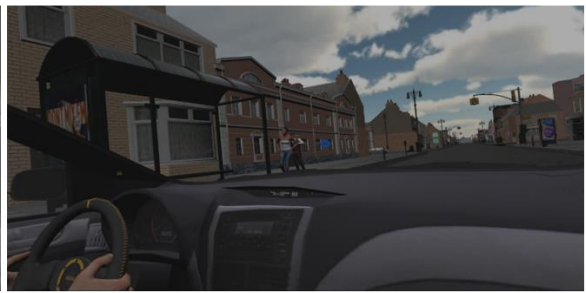

### 2.3.1.2 Kitchen: Multitasking and prospective memory task

In the kitchen, (i.e., scene 6; see Table 1), the user should complete two main tasks: the cooking task (i.e., preparing breakfast) and a prospective memory task. The cooking task encompasses frying an omelet and sausages, putting a chocolate pie in the oven, as well as boiling some water for a cup of tea or coffee. The user must handle two pans (one for the omelet and one for the sausages) and a kettle. Images of the omelet and sausages are presented above the cooker to display what their appearance should be when they are ready. Scoring relies on the animations from each game object (i.e., the omelet and the sausages). At the beginning of the animation, both items have a reddish (raw) color which gradually turns to either a yellowish (omelet) or brownish (sausages) color, and finally both turn to black (burnt). The score for each pan hence depends on the time that the user removes the pans from the stove (pauses/stops the animation) and places them on the kitchen worktop (for calculation of the score, see Figure 4). Equally, the score for boiling the kettle is measured in relation to the stage of the audio playback (e.g., the water is ready when the kettle whistles; see Figure 4) when the kettle is placed on the kitchen worktop.

After breakfast, the user needs to take his/her meds (i.e., a prospective memory task). When the user has had his/her breakfast, the final button of the scene appears (see Figure 3). The user should press this button to confirm that all the tasks in the scene are completed. If the user has already taken his/her medication before pressing the final button, then the scene ends, and the user receives 6 points. Otherwise, the first prompt appears (i.e., "You Have to Do Something Else"). If the user then follows the prompt and takes their medication, they receive 4 points. If the user presses the final button again, then the second prompt appears (i.e., "You Have to Do Something After Having your Breakfast"). If the user follows this prompt and takes their medication, they receive 2 points. If the user presses the final button again, then the third prompt appears (i.e., "You Have to Take Your Meds"). If the user follows this prompt and takes their medication, they then receive 1 point. If the user represses the final button without ever taking their medication, they get zero points, and the scene ends.

Figure 4. A Schematic Representation of the Cooking Task scoring.

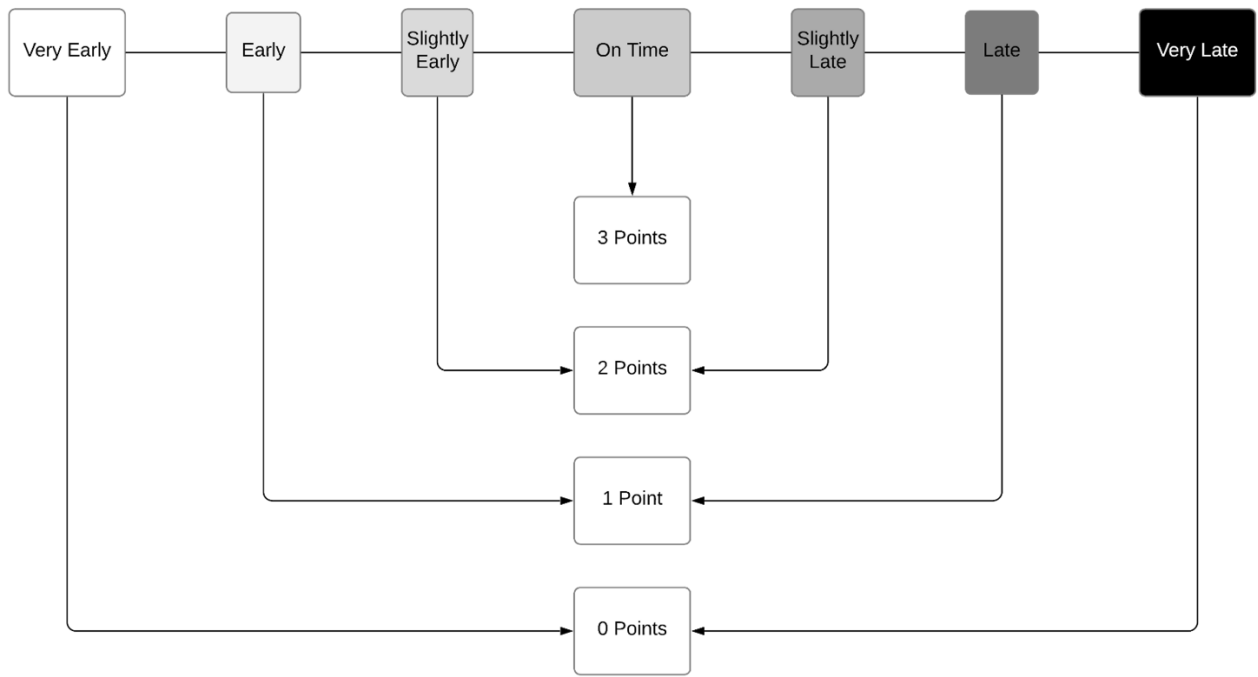

| Cooking Item | Very Early | Early | Slightly Early | On Time | Slightly Late | Late | Very Late |
|---|---|---|---|---|---|---|---|
| Pan-Omelette | 0-13.99 secs | 14-15.99 secs | 16-17.99 secs | 18-22 secs | 22.01-23.99 secs | 24-26 secs | > 26 secs |
| Pan-Sausages | 0-17.99 secs | 18-19.99 secs | 20-21.99 secs | 22-26 secs | 26.01-27.99 secs | 28-30 secs | > 30 secs |
| Kettle | 0-10.99 secs | 11-12.99 secs | 13-14.99 secs | 15-17 secs | 17.01-18.99 secs | 19-21 secs | > 21 secs |

### 2.3.1.3 Living room: Selective visuospatial attention and prospective memory task

In the living room (i.e., scene 8; see Table 1), the user should collect six items (i.e., a red book, £20, a smartphone, a library card, the flat keys, and the car keys) that are placed in various locations within the living room (see Figure 3). The user is not required to memorize the items since there is a reminder list on one of the walls of the living room. The user collects the items by attaching them to the snap-drop-zones attached to the left controller. The user receives 1 point for each item collected. However, there are distractors (e.g., magazines, books, a remote control, a notebook, a pencil, a chessboard, and a bottle of wine) in the room. If the user attempts to collect one of the distractors, the item falls (only the target items can be attached to the snap-drop-zones), which counts as an error. After collecting all the objects, the user needs to take the chocolate pie out of the oven and place it on the kitchen worktop before leaving the apartment (prospective memory task; see scoring for medication above).

### 2.3.1.4 Garden: Prospective memory task

In the garden (i.e., scene 10; see Table 1), the user initiates a conversation with Alex (an NPC), to perform a distractor task (i.e., to open the gate). The conversation continues after this distractor task, where the user needs to respond to Alex's question (i.e., "Do we need to do something else at this time?") by selecting either "yes" or "no" (see Figure 3). This action is considered as the first prompt for the prospective memory task, and if the user responds "yes", then the second interactive board appears (see Figure 5 for scoring). If the user selects "no", then the second prompt is given by Alex (i.e., "Are you sure that we do not have to do something around this time?"). If the user selects "no", then the third prompt is provided by Alex (i.e., "I think that we have to do something around this time."). If the user again selects "no", clarification is provided by Alex (i.e., "Oh yes, we need to call Rose"), and the user receives 0 points (see Figure 5).

When the user chooses "yes", the second interactive board appears. This second interactive board displays eight items (see Figure 3). There is one item, which presents the correct prospective memory response (i.e., the smartphone). There are also three items which are responses related to the other prospective memory tasks (i.e., a red book, carrot cake, flat keys). There is one item, which is semantically related to the correct prospective memory response (i.e., a tablet computer). Also, there are three items which are unrelated distractors, which are neither related to the other prospective memory tasks, nor are in the same semantic category as the correct prospective memory response (i.e., ice cream and a smartphone). Scoring depends on the user's responses on the first and the second interactive boards (see Figure 5).

Figure 5. Prospective Memory: Positive Scoring System

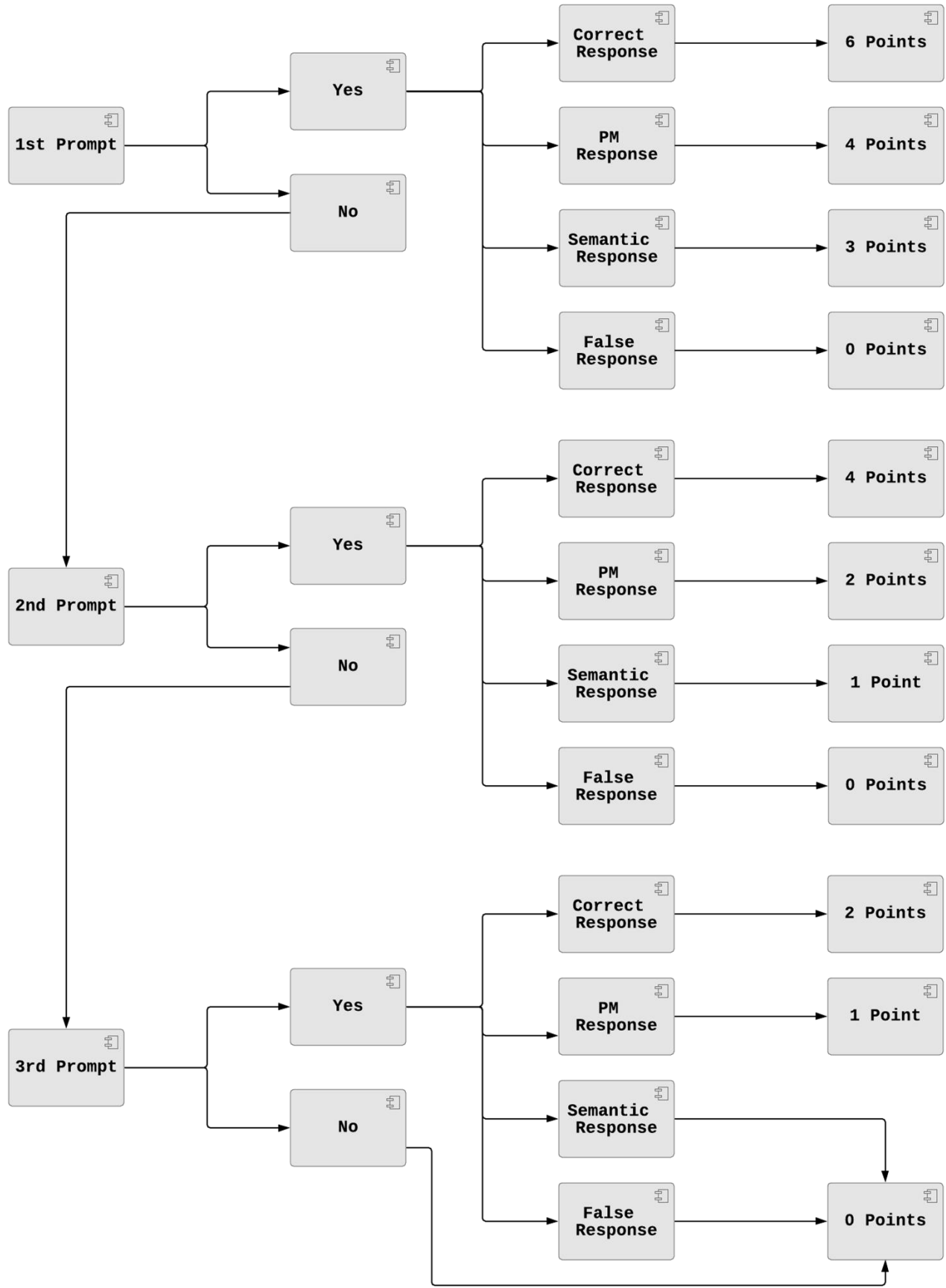

### 2.3.2 In the city

#### 2.3.2.1 On the road: Selective visual attention

In this scenario, the user is a passenger in a car with Alex driving (i.e., scene 12; see Table 1). The radio is on, and the speaker announces a competition where the user needs to identify all the radio stations' target posters and avoid the distractor ones (see Figure 3), which are hung along the street. There are 16 target posters and 16 distractors equally allocated on both sides of the street. Eight of the distractors have the same shape as the target poster, but a different background color. The other eight distractors have the same background color as the target posters, but they are a different shape (see Figure 3). The user is awarded 1 point when a target poster is "spotted" and subtracted 1 point when a distractor poster is "spotted". The maximum score is 16, and the number of correctly identified posters and distractors (for each type) identified on each side of the road is recorded.

#### 2.3.2.2 Supermarket: Delayed recognition and prospective memory task

The user arrives at the supermarket (i.e., scene 14; see Table 1), where s/he should buy the items from the shopping list. The user navigates within the shop by following the arrows, and collects the items using the snap-drop-zones attached to the left controller (see Figure 6). The items on the shelves of the supermarket are the same items as the immediate recognition task in scene 3 (see Bedroom). The scoring system is identical to the immediate recognition task in scene 3 (see Bedroom), and the score is calculated when the user arrives at the till to buy the items. Outside the supermarket (i.e., scene 15), the user has another conversation with Alex, where s/he needs to remember that they must collect the carrot cake at 12 noon (i.e., a prospective memory task). The conversation is performed and scored in the same way as the prospective memory task in scene 10 (see Garden). The user then goes with Alex to the bakery to collect the carrot cake.

#### 2.3.2.3 Bakery and library: prospective memory tasks

The user is outside the bakery (i.e., scene 16; see Table 1), after already collecting the carrot cake. Here, they have another interaction with Alex where he asks, "Do we need to do something else at this time?" However, this time, there is no prospective memory task to perform and the user should respond negatively. This deception helps to examine whether the user is simply responding affirmatively to all prospective memory task prompts. If the user responds affirmatively (i.e., "yes"), then s/he loses points (see Figure 7). This conversation is similar to the prospective memory task in scene 10 (see Garden). However, the scoring is now inverted, where the user should choose "no" three times in response to Alex's prompts to avoid points being subtracted. In the prospective memory task that follows in the next scene, the user should again respond negatively to avoid a maximum of 3 points being deducted (see Figure 7). Therefore, in this task, up to 6 points may be subtracted. Then, the user arrives at the library (i.e., scene 17; see Table 1), where s/he has another interaction with Alex (i.e., a prospective memory task), which is performed and scored in the same way as the prospective memory task in scene 10 (see Garden and Figure 5). After leaving the library, Alex and the user proceed to the petrol station to refill the car.

Figure 6. VR-EAL Storyline: Scenes 14 - 22

**Scene 14**  **Scene 14**

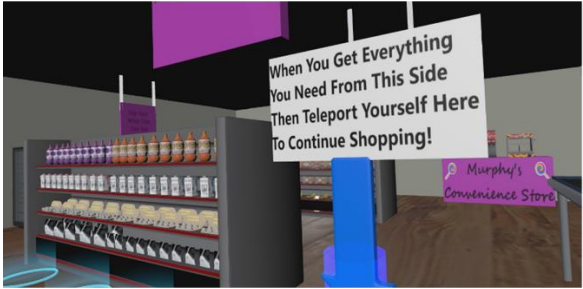 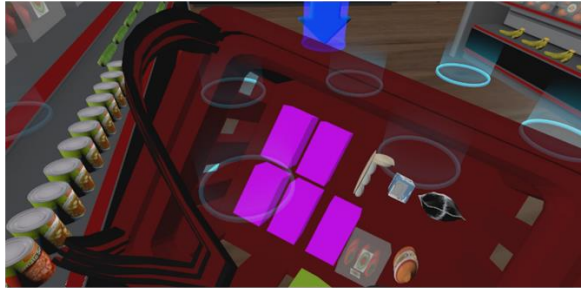

**Scene 15**  **Scene 17**

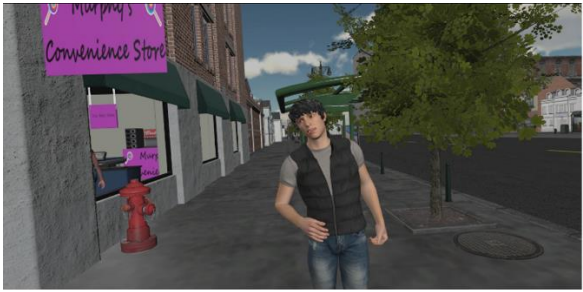 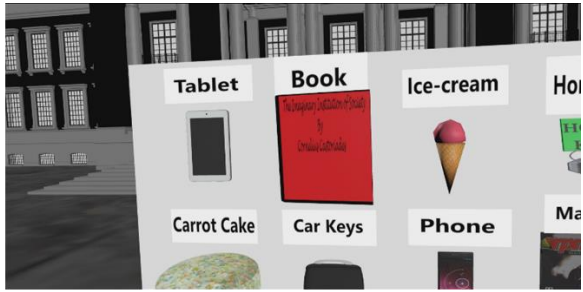

**Scene 19**  **Scene 19**

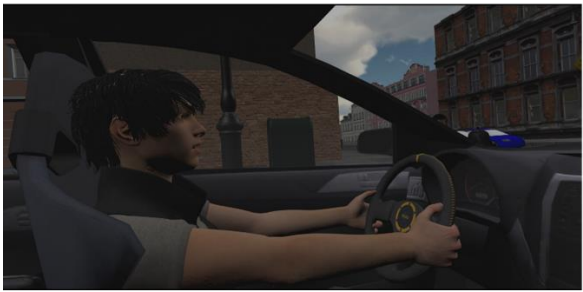 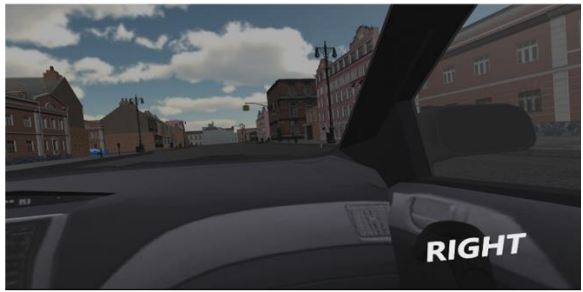

**Scene 20**  **Scene 22**

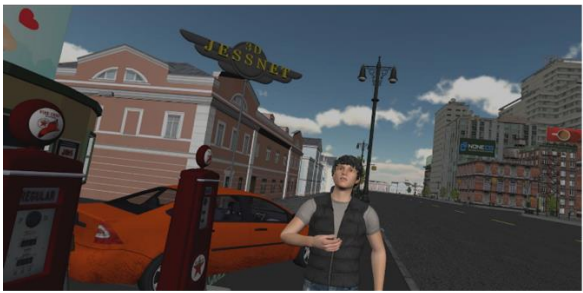 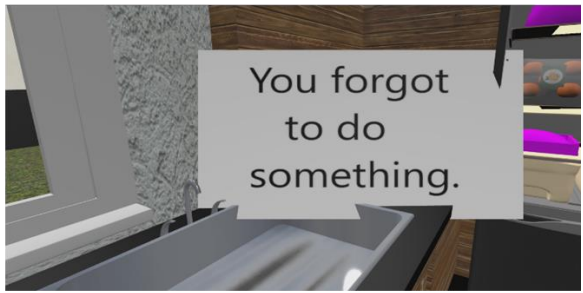

**Scene 22**  **Scene 22**

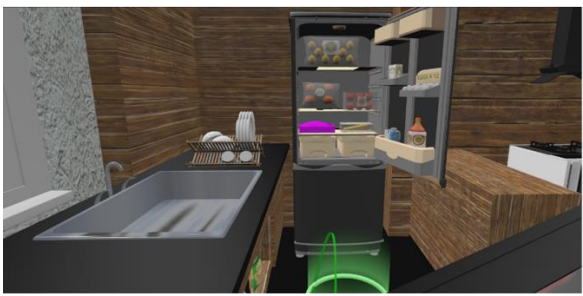 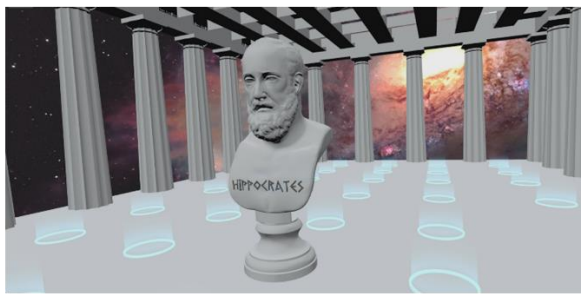

Figure 7. Prospective Memory: Negative Scoring System

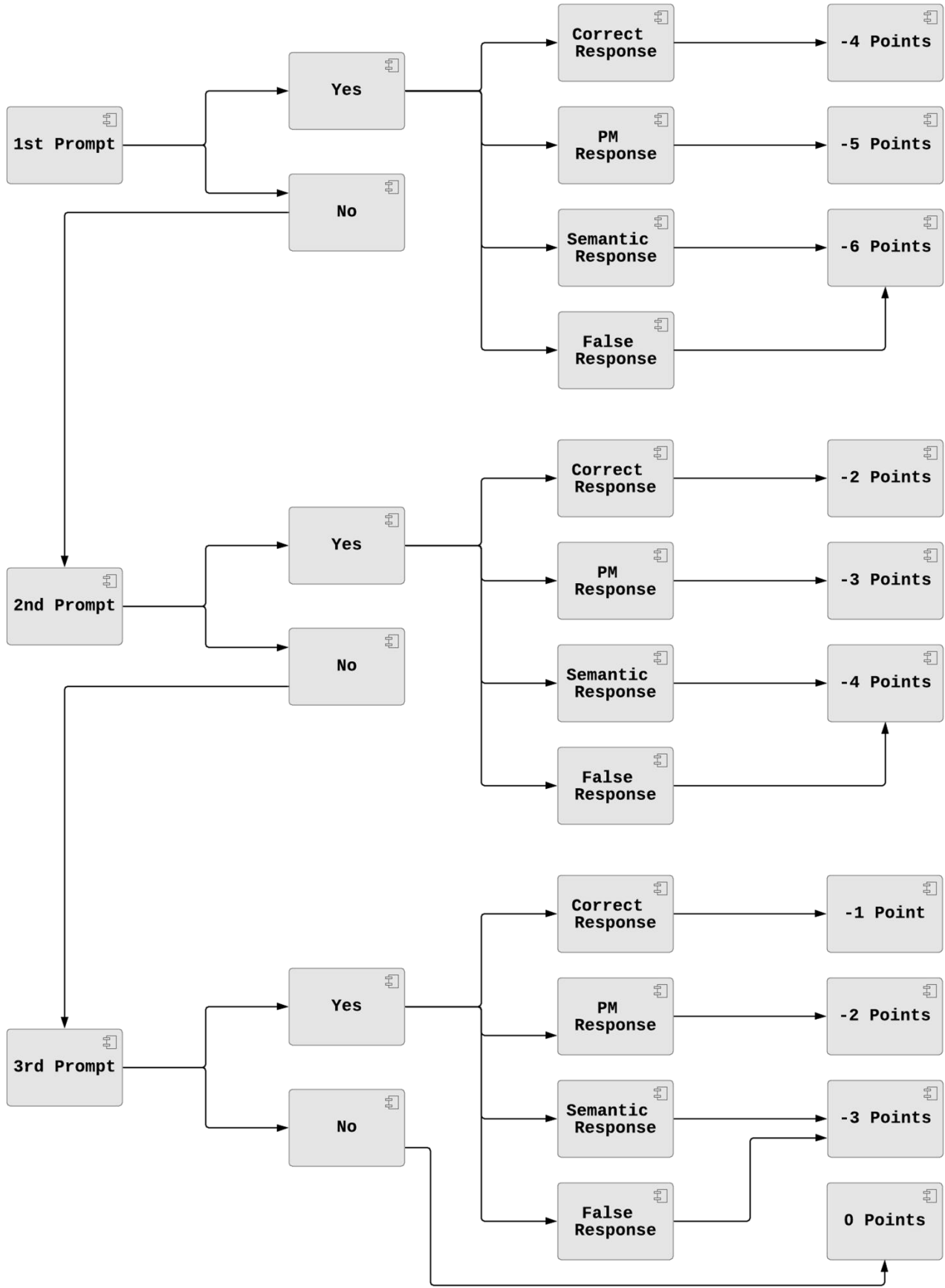

### 2.3.2.4 On the road home: Selective auditory attention and prospective memory tasks

The user is in the car with Alex and the radio station has another challenge (i.e., scene 19; see Table 1 and Figure 6). This time small speakers playing different sounds have been placed on both sides of the street. The user should detect the target sounds and avoid the false high-pitched and low-pitched sounds while Alex drives along the street. As in the tutorial, the user presses the controller trigger when they detect a sound. If the user presses the trigger on the right controller to detect a target sound originating on the right side, then s/he gets 2 points. If the user presses the trigger on the left controller to detect a target sound originating on the left side, s/he also gets 2 points. If the user presses the trigger on the right controller to detect a target sound originating on the left side or a trigger on the left controller to detect a target sound originating on the right side, s/he gains only 1 point. On the other hand, if the user responds to a distractor sound, irrelevant of its origin or the controller used to respond, 1 point is deducted. The stored data summarize the number of detected sounds of each type (i.e., target sounds, low pitched distractor sounds, high pitched distractor sounds), the number of sounds detected on the left and right sides, and how many times the wrong controller (i.e., false side) was used to detect a target sound.

After the car ride, the user is at the petrol station with Alex (i.e., scene 20; see Table 1). The user has another conversation with Alex, where s/he receives false prompts (i.e., there is not a prospective memory task to perform). This prospective memory task is performed and scored in the same way as the Bakery prospective memory task (i.e., scene 20, see Figure 7). Then, the user returns back home with Alex (i.e., scene 21), where the user has their last interaction with Alex, and should give him the extra pair of keys to the flat. This prospective memory task is also performed and scored as the prospective memory task in scene 10 (i.e., see Garden and Figure 5).

### 2.3.2.5 Back Home: Distractor and prospective memory task

In the final scene (i.e., scene 22), the user is back home, where s/he is required to perform two tasks (see Figure 6). The first task is a distractor task, where the user needs to put away the items that s/he has bought from the supermarket. While doing this, s/he needs to remember that s/he should take his/her medication at 1pm. If the user performs the task on time, then s/he receives 6 points. If the user fails to remember the prospective memory task after 70 seconds, a prompt appears. If the user performs the task after this first prompt, s/he receives 4 points. If the user ignores the first prompt, after another 10 seconds, a second prompt appears. If the user performs the task after the second prompt, s/he receives 2 points. If the user ignores the second prompt, after a further 10 seconds, a third and final prompt appears. If the user performs the task after the third prompt, s/he receives 1 point. If the user ignores the third prompt and presses the final button, s/he receives 0 points.

Once the user presses the final button, the scenario finishes and the credits appear. Here, the user is informed of the aims of VR-EAL. The VR-EAL attempts to be an extensive neuropsychological assessment of prospective memory, episodic memory, executive functions, and attentional processes by collecting various data pertinent to these cognitive functions (see Supplementary Material I for an example of VR-EAL data).

### 2.4 Development of VR software in Unity

The scenario provides the main framework for developing the VR application. VR-EAL was developed to be compatible with the HTC Vive, HTC Vive Pro, Oculus Rift, and Oculus Rift-S to be aligned with the minimum hardware technological specifications for guaranteeing health and safety standards and the reliability of the data (Kourtesis *et al.*, 2019a). The quality of VR-EAL was assessed in terms of user experience, game mechanics, in-game assistance, and VRISE using the Virtual Reality Neuroscience Questionnaire (VRNQ; Kourtesis *et al.*, 2019b). The total duration for the VR neuropsychological assessment is approximately 60 minutes, which falls within the suggested maximum duration range for VR sessions (Kourtesis *et al.*, 2019b). Long VR sessions appear more susceptible to VRISE, though, long (50-70 minutes) VR sessions which exceed the parsimonious cut-offs from the VRNQ do not induce VRISE (Kourtesis *et al.*, 2019b). For this reason, the parsimonious cut-offs for the VRNQ (see Table 2) will be used to ensure that VR-EAL users do not suffer from VRISE (Kourtesis *et al.*, 2019b).

Table 2. VRNQ minimum and parsimonious cut-offs

| Score | Minimum Cut-offs | Parsimonious Cut-offs |
|---|---|---|
| User Experience | $\geq 25/35$ | $\geq 30/35$ |
| Game Mechanics | $\geq 25/35$ | $\geq 30/35$ |
| In-Game Assistance | $\geq 25/35$ | $\geq 30/35$ |
| VRISE | $\geq 25/35$ | $\geq 30/35$ |
| VRNQ Total Score | $\geq 100/140$ | $\geq 120/140$ |

*The median of each sub-score and total score should meet the suggested cut-offs to determine that the evaluated VR software is of adequate quality without any significant VRISE. The utilization of the parsimonious cut-offs more robustly supports the suitability of the VR software. Derived from Kourtesis et al. (2019b)*

The development of VR-EAL should be proximal to commercial VR applications. The first step of the development is to select Unity's settings to support the development of VR software. For the development of VR-EAL, Unity version 2017.4.8f1 was used. Unity supports VR software development kits (SDK). The built-in support for the SDKs is for the OpenVR SDK and the Oculus SDK. In the player settings of Unity, the developer may select the VR/XR supported box, which allows the addition of the aforementioned SDKs. For VR-EAL, Unity's support for both the OpenVR SDK and the Oculus SDK were added, though, priority was given to the OpenVR SDK.

### 2.4.1 Navigation and interactions

VR software for the cognitive sciences may require intensive movement and interactions. However, the development of such interactions demands highly advanced programming skills in C# and expertise in VR software development in Unity. Nonetheless, on Unity's asset store and GitHub's website, there are some effective alternatives that facilitate the implementation of intensive interactions without the requirement of highly advanced software development skills. The utilization of the SteamVR SDK, Oculus SDK, Virtual Reality Toolkit (VRTK) or similar toolkits and assets are options which should be considered. For the development of VR-EAL, the SteamVR SDK and VRTK were selected to develop accurate interactions compatible with the capabilities of the 6DoF controllers of HTC Vive and Oculus Rift. The advantage of SteamVR SDK, which was developed based on OpenVR SDK, is that is compatible with both the HTC Vive and Oculus Rift, though, it does not offer a wide variety of interactions or good quality physics. Nonetheless, the VRTK mounts the SteamVR SDK and offers better quality physics and plenty of interactions that support the development of VR research software for cognitive sciences.

A fundamental interaction in the VE is navigation. HTC Vive and Oculus Rift offer a play area of an acceptable size, which permits ecologically valid scenarios and interactions to be developed (Porcino *et al.*, 2017; Borrego *et al.*, 2018). However, the VR play area is restricted to the limits of the physical space and tracking area; hence, it does not allow navigation which is based on physically walking (Porcino *et al.*, 2017; Borrego *et al.*, 2018). A suitable solution is the implementation of a navigation system based on teleportation. Teleportation enables navigation exceeding the boundaries of the VR play area and delivers high-level immersion, a pleasant user experience, and decreases the frequency of VRISE. Typically, a navigation system of a VR software which depends on a touchpad, keyboard, or joystick, substantially increasing the frequency and intensity of VRISE (Bozgeyikli *et al.*, 2016; Frommel *et al.*, 2017; Porcino *et al.*, 2017). In VR-EAL, a combination of teleportation and physical movement (i.e., free movement of the upper limbs and walking in a small-restricted area) is used (see Figures 1, 2, 3, and 6).

The VRTK provides scripts and tools that aid the developer to build a teleportation system. The VRTK is compatible with 6DoF controllers, which are necessary to provide naturalistic and ergonomic interactions. In addition, the implementation of 6DoF controllers facilitates familiarization with their controls and their utilization, because they imitate real life hand actions and movements (Sportillo *et al.*, 2017; Figueiredo *et al.*, 2018). The VR-EAL user learns the controls in the tutorials, though, there are also in-game instructions and aids that assist even a non-gamer user to grab, use, and manipulate items. These in-game assistance methods significantly alleviate the occurrence of VRISE, while increasing the user's level of enjoyment (Caputo *et al.*, 2017; Porcino *et al.*, 2017). Finally, the VRTK offers additional gamified interactions through the snap-drop-zones. The snap-drop-zones are essentially carriers of game objects and their mechanics are similar to the trigger-zones. For example, when a game object (i.e., a child object of a controller) enters the zone, if the game object is released (i.e., stops being a child object of the controller), this game object is attached to the snap-drop-zone (i.e., it becomes a child object). In VR-EAL, the snap-drop-zones are extensively used, allowing the scoring of tasks, which otherwise would be less effective in terms of accuracy of response times.

The interaction and navigation systems are essential to increase immersion. However, immersion depends on the strength of the placement, plausibility, and embodiment illusions (Slater, 2009; Slater *et al.*, 2010; Maister *et al.*, 2015; Pan & Hamilton, 2018). An ecologically valid neuropsychological assessment necessitates genuine responses from the user. Robust placement and plausibility illusions ensure that the user will genuinely perform the tasks as s/he would perform them in real life (Slater, 2009; Slater *et al.*, 2010; Pan & Hamilton, 2018). The placement illusion is the deception of the user that s/he is in a real environment and not in a VE (Slater, 2009; Slater *et al.*, 2010). However, the placement illusion is fragile because the VE should react to the user's actions (Slater, 2009; Slater *et al.*, 2010). This is resolved by the plausibility illusion, which is the deception of the user that the environment reacts to his/her actions. Therefore, the user believes the plausibility of being in a real environment (Slater, 2009; Slater *et al.*, 2010). The naturalistic interactions in the VE that VRTK and SteamVR SDK offer are pertinent to the plausibility illusion.

### 2.4.2 Graphics

A strong placement illusion relies on the quality of the graphics and 3D objects (Slater, 2009; Slater *et al.*, 2010). Correspondingly, the quality of the graphics principally depends on the rendering (Lavoué & Mantiuk, 2015). The rendering comprises the in-game quality of the image (i.e., perceptual quality), and the omission of unnecessary visual information (i.e., occlusion culling) (Lavoué & Mantiuk, 2015). The advancement of these rendering aspects ameliorates both the quality of graphics and the performance of the VR software (Brennesholtz, 2018). Likewise, the amplified image refresh rate and resolution decrease the frequency and intensity of VRISE (Brennesholtz, 2018). However, the rendering pipeline and shaders in Unity are not optimized to meet VR standards.

The VR software developer should select different rendering options, so the quality of graphics is good and the image's refresh rate is equal to or above 90Hz, which is the minimum for high-end HMDs like the HTC Vive and Oculus Rift. For example, the "Lab renderer" is an asset that allows VR optimized rendering and replaces the common shaders with VR optimized ones. Additionally, the "Lab renderer" supports an extensive number of light sources (i.e., up to 15), which otherwise would not be feasible in VR. However, the developer needs to build a global illumination map (i.e., lightmap), which substantially alleviates the cost of lights and shadows on the software's performance (Jerald, 2015; LaViola et al., 2017; Sherman & Craig, 2018). Usually, the lightmapping process is the final step in the development process.

The acquisition of 3D objects may be expensive or time-consuming. However, there are several free 3D objects on Unity's asset store and webpages, such as TurboSquid and Cgtrader, which can be used for the development of VR research software. Importantly, the license for these 3D objects obliges the developer not to use them for commercial purposes. However, research VR software like VR-EAL is free, and research software developers usually do not commercialize their products. Although there are several free 3D objects on the websites mentioned above, it is likely that these 3D objects are not compatible with VR standards. In VR, the 3D objects should comprise a low number of polygons (Jerald, 2015; LaViola et al., 2017; Sherman & Craig, 2018). A decrease in polygons may be achieved using software like 3DS Max. The optimization of the 3D objects (to meet VR standards) may be achieved by simply importing the 3D objects, optimizing them, and then exporting them with a low number of polygons in a Unity compatible format (i.e., fbx and obj).

Nevertheless, developers often aim to create large VEs such as cities, towns, shops, and neighborhoods. Each 3D object, whether it be small (e.g., a pen), medium (e.g., a chair), or large (e.g., a building), may comprise several mesh renderers. Unity requires one batch (i.e., draw call) for each mesh renderer. In large environments, the batching may significantly lower the image's refresh rate and the overall performance of the software (Jerald, 2015; LaViola et al., 2017; Sherman & Craig, 2018). However, assets like MeshBaker are designed to solve this problem. MeshBaker merges all the selected textures and meshes into a clone game object with a small number of meshes and textures. For example, the town that was designed for VR-EAL required >1000 draw calls. After the implementation of MeshBaker, the draw calls were decreased to 16. However, the disadvantage of MeshBaker is that it does not clone the colliders. Hence, the developer needs to deactivate the mesh renderers of the original game object(s) and leave active all the colliders, while the original game object(s) should be precisely in the same position with the clone(s) so the colliders of the former are aligned with the meshes of the latter. Of note, MeshBaker should be purchased from Unity's asset store in contrast with the other assets used in VR-EAL's development which are freely available (i.e., SteamVR SDK, VRTK, and Lab renderer).

### 2.4.3 Sound

Another important aspect of VR software development is the quality of the sound. The addition of spatialized sounds in the VE (e.g., ambient and feedback sounds) augments the level of immersion and enjoyment (Vorländer & Shinn-Cunningham, 2014), and significantly reduces the frequency of VRISE (Viirre *et al.*, 2014). Spatialised sounds in VR assist the user to orient and navigate (Ferrand *et al.*, 2017), and enhance the geometry of the VE without reducing the software's performance (Kobayashi *et al.*, 2015). In Unity, a developer may use tools like SteamAudio, Oculus Audio Spatializer, or Microsoft Audio Spatializer for good quality and spatialization of the audio aspects. In VR-EAL's development, Steam Audio was used. SteamAudio spatializes the sound to the location of the audio source's location and improves the reverberance of sounds (i.e., Unity's reverb zone).

Notably, the strength of the plausibility illusion is analogous to the sensorimotor contingency, which is the integration of the senses (i.e., motion, vision, touch, smell, taste) (Gonzalez-Franco & Lanier, 2017). Moreover, the VRTK enables the utilization of a haptic modality. For example, when the user grabs an item in the VE, s/he expects a tactile sense as would be experienced in real life. The haptic feedback of the VRTK allows the developer to activate/deactivate the vibration system of the 6DoF controllers when an event occurs (e.g., grabbing or releasing a game object) and define the strength and the duration of the vibration. The spatialized audio and the haptics additionally reinforce the plausibility illusion by providing an expected auditory and haptic feedback to the user (Jerald, 2015; LaViola et al., 2017; Sherman & Craig, 2018).

### 2.4.4 3D characters

Furthermore, VR research software like VR-EAL, which includes social interactions with virtual characters, should also consider the quality of the 3D characters in terms of realistic appearance and behavior. For example, Morph 3D and Mixamo both offer free and low-cost realistic 3D characters that may be used in VR software development. For VR-EAL, Morph 3D was preferred, though, other virtual humans from Unity's asset store were used to populate the scenes (e.g., individuals waiting for the bus at the bus stop). The 3D characters provided by Morph 3D have modifiable features, which may be used by the developer to customize the character's appearance (e.g., body size) and expressions (e.g., facial expressions which signify emotions such as happiness and sadness). Morph 3D provides two free 3D characters (i.e., female and male) capable of displaying naturalistic behavior (i.e., body and facial animations). The developer may use body animations which derive from motion capture (MoCap) techniques. For the development of VR-EAL, body animations were derived from free sample animations from Unity's Asset Store (e.g., hand movement during talking, and waving) and the MoCap animations library of the Carnegie Mellon University. However, the effective implementation of the animations requires modification and synchronization (e.g., the animation should be adjusted to the length of the 3D character's interaction) using Unity's animation and the animator's windows. The animation window may be used for synchronization, while the animator is a state machine controller that controls the transition between animations (e.g., when this event happens, play this animation, or when animation X ends, play animation Y).

However, the most challenging aspect of realistic 3D characters is the animation of their facial features. The 3D character should have realistic eye interactions (i.e., blinking, looking at or away from the user) and talking (i.e., a realistic voice and synchronized lip movements). Limitations in both time and resources did not allow for seamless face and body animations since that would require multimillion dollars' worth of equipment like those used by big game studios. This limitation can result in an uncanny valley effect (Mori *et al.*, 2012; Seyama & Nagayama, 2007). However, previous research has shown that, when users interact with 3D humanoid embodied agents that have the role of an instructor (like the ones used in VR-EAL), they have less expectations for that character due to their role and limited interactivity (Korre, 2019). The addition of 3D characters was important because they deliver an interaction metaphor resembling human-to-human interactions (Korre, 2019). Even though adding a 3D character in the scene can introduce biases, the illusion of humanness - which is defined as the user's notion that the system (in this case the 3D NPC) possesses human attributes and/or cognitive functions - has been found to increase usability (Korre, 2019).

Realistic voices may be established by employing voice actors to produce the script. However, the employment of temporary staff increases development costs. For VR-EAL, text-to-speech technologies were used as an alternative solution to deliver realistic voices. Balabolka software was used in conjunction with Ivona3D Voices (n.b., Ivona3D has been replaced by Amazon Polly). Balabolka is an IDE for text-to-speech which allows further manipulation of voices (i.e., pitch, rate, and volume), while Ivona3D provides realistic voices. The developer types or pastes the text into Balabolka, Balabolka modifies it with respect to the desired outcome (e.g., high-pitched or low-pitched voice) and exports the file in a .wav format. Additionally, free software like Audacity may be used, which offers greater variety in sound modifications. The second crucial part is to synchronize the eyes and lip movements with the voice clips and body animations. There are assets on Unity's asset store that may be used to achieve this desired outcome. In VR-EAL, Salsa3D and RandomEyes3D were used to attain good quality facial animations and lip synchronization. Salsa3D synchronizes the lips with the voice clip, while RandomEyes3D allows the developer to control the proportion of eye contact with the user for each voice clip.

### 2.4.5 Summary of the VR-EAL illusions

Summing up, the described VR-EAL development process facilitates the utilization of ergonomic interactions, a VR compatible navigation system, good quality graphics, haptics, and sound, as well as social interactions with realistic 3D characters. These software features contribute to the lessening or avoidance of VRISE and augmentation of the level of immersion by providing placement and plausibility illusions. However, VR-EAL does not seem to deliver a strong embodiment illusion (i.e., the deception that the user owns the body of the virtual avatar), because it only relies on the presence of the 6DoF controllers. A possible solution would be the implementation of inverse kinematics, which animates the virtual avatar with respect to the user's movements. In addition, the temporal illusion (i.e., deceiving the user into thinking that the virtual time is real-time) only relies on changes in environmental cues (e.g., the movement of the sun, and changes in lighting). Therefore, a VR digital watch was developed (freely distributed on GitHub) and used in an attempt to increase the strength of the temporal illusion. To conclude, the development of VR research software is feasible mainly using free or low-cost assets from GitHub, Unity Asset's store, and other webpages. However, the suitability and quality of the VR software should be evaluated before its implementation in research settings.

## 3    Evaluation of VR-EAL

### 3.1    Participants

Twenty-five participants (6 female gamers, six male gamers, seven female non-gamers, and six male non-gamers) were recruited for the study via the internal email network of University of Edinburgh as well as social media. The mean age of the participants was 30.80 years (SD = 5.56, range = 20-45) and the mean years of full-time education was 14.20 years (SD = 1.60, range = 12-16).  Twelve participants (3 female gamers, 3 male gamers, 3 female non-gamers, and 3 male non-gamers; mean age = 30.67 years, SD = 2.87, range = 26-36; mean educational level = 14.75 years, SD = 1.30, range = 12-16 years) attended all three VR sessions (i.e., alpha, beta, and final versions), while the remaining 13 participants only attended the final version session. The gamer experience was a dichotomous variable (i.e., gamer or non-gamer) based on the participants' response to a question asking whether they played games on a weekly basis. The current study has been approved by the Philosophy, Psychology and Language Sciences Research Ethics Committee of the University of Edinburgh. All participants were informed about the procedures, possible adverse effects (e.g., VRISE), data utilization, and the general aims of the study both orally and in writing; subsequently, every participant gave written informed consent.

### 3.2 Material

#### 3.2.1 Hardware and software

An HTC Vive HMD, two lighthouse-stations for motion tracking, and two 6DoF controllers were used. The HMD was connected to a laptop with a 2.80GHz Intel Core i7 7700HQ processor, 16 GB RAM, a 4095MB NVIDIA GeForce GTX 1070 graphics card, a 931 GB TOSHIBA MQ01ABD100 (SATA) hard disk, and Realtek High Definition Audio. The size of the VR play area was 4.4 m2. The software was the alpha version of VR-EAL for session 1, the beta version of VR-EAL for session 2, and the final version of VR-EAL for session 3.

#### 3.2.2 VRNQ

The VRNQ is a paper-and-pencil questionnaire containing 20 questions, where each question refers to one of the criteria necessary to assess VR research/clinical software in neuroscience (Kourtesis *et al.*, 2019b). The 20 questions assess four domains: user experience, game mechanics, in-game assistance and VRISE. The VRNQ has a maximum total score of 140, and 35 for each domain. VRNQ responses are indicated on a 7-point Likert style scale ranging from 1= extremely low to 7 = extremely high. Higher scores indicate a more positive outcome; this also applies to the evaluation of VRISE intensity. Hence, higher VRISE scores indicate lower intensities of VRISE (i.e., 1 = extremely intense feeling, 2 = very intense feeling, 3 = intense feeling, 4 = moderate feeling, 5 = mild feeling, 6 = very mild feeling, 7 = absent). Additionally, the VRNQ allows participants to provide qualitative feedback, which may be useful during the development process. Lastly, the VRNQ has two cut-off scores, the minimum (i.e., 25 for every sub-score, and 100 for the total score) and parsimonious (i.e., 30 for every sub-score, and 120 for the total score) cut-offs. The median scores derived from the user sample should exceed at least the minimum cut-offs, while for VR software which requires long VR sessions, then the parsimonious cut-offs should be preferred. For the evaluation of VR-EAL, the parsimonious cut-offs were opted to support the suitability of VR-EAL. The VRNQ can be downloaded from Supplementary Material II.

### 3.3 Procedures

Twelve participants attended all three VR sessions, while an additional 13 participants only attended the third session. The period between each session was six to eight weeks. In each session, participants were immersed in a different version of VR-EAL. Each session began with inductions in VR-EAL, the HTC Vive, and the 6DoF controller. Then, participants played a version of VR-EAL. Lastly, after the completion of VR-EAL, participants were asked to complete the VRNQ. A preview of the final version of VR-EAL can be found in Supplementary Material III or by following the hyperlink: https://www.youtube.com/watch?v=IHEIvS37Xy8&t= .

### 3.4 Statistical Analysis

Bayesian statistics were preferred over null hypothesis significance testing (NHST). P-values calculate the distance (i.e., the difference) between the data and the null hypothesis (H0) (Cox & Donnelly, 2011; Held & Ott, 2018). The p-values assess the assumption of no difference or no effect, while the Bayesian factor ($BF_{10}$) converts p-values into evidence in favor of the alternative hypothesis (H1) against the H0 (Cox & Donnelly, 2011; Held & Ott, 2018). $BF_{10}$ is found robustly more parsimonious than the p-value in evaluating the evidence against the H0 (Cox & Donnelly, 2011; Held & Ott, 2018; Wagenmakers et al., 2018a; Wagenmakers et al., 2018b). Importantly, the difference between $BF_{10}$ and p-values is even greater (in favor of $BF_{10}$) in small sample sizes, where $BF_{10}$ should be opted for as it is more parsimonious (Held & Ott, 2018; Wagenmakers *et al.*, 2018a;

Wagenmakers *et al.*, 2018b). For these reasons, the $BF_{10}$ was preferred instead of p-values for the assessment of statistical inference, especially while having a relatively small sample size. Moreover, a larger $BF_{10}$ postulates more evidence in support of H1 (Cox & Donnelly, 2011; Held & Ott, 2018; Marsman & Wagenmakers, 2017; Wagenmakers *et al.*, 2018a; Wagenmakers *et al.*, 2018b). Specifically, a $BF_{10} \le 1$ indicates no evidence in favor of H1, while $1 < BF_{10} < 3$ indicates anecdotal evidence for H1, $3 \le BF_{10} < 10$ indicates moderate evidence for H1, $10 \le BF_{10} < 30$ indicates strong evidence for H1, $30 \le BF_{10} < 100$ indicates very strong evidence for H1, and a $BF_{10} \ge 100$ indicates extreme evidence for H1 (Marsman & Wagenmakers, 2017; Wagenmakers *et al.*, 2018a; Wagenmakers *et al.*, 2018b). For our analyses, we accept the notion put forward by Marsman & Wagenmakers, (2017), Wagenmakers *et al.*, (2018a), and Wagenmakers *et al.*, (2018b) of $BF_{10} \le 1$ indicating no evidence in favor of H1, $BF_{10} > 3$ indicating moderate evidence in favor of H1, $BF_{10} \ge 10$ indicating strong evidence for H1, and $BF_{10} \ge 100$ indicating extreme evidence for H1. In this study, a parsimonious threshold of $BF_{10} \ge 10$ was set for statistical inference, which postulates strong evidence in favor of the H1 (Marsman & Wagenmakers, 2017; Wagenmakers *et al.*, 2018a; Wagenmakers *et al.*, 2018b), and corresponds to a p-value <.01 (e.g., $BF_{10} = 10$) or to a p-value <.001 (e.g., $BF_{10} > 11$) (Cox & Donnelly, 2011; Held & Ott, 2018). However, we report both $BF_{10}$ and p-values in this study. A Bayesian paired samples t-test was performed to compare the VRNQ results for each version of VR-EAL (N=12), as well as to inspect potential differences between gamers (N = 12) and non-gamers (N = 13). The Bayesian statistical analyses were performed using JASP (Version 0.8.1.2) (JASP Team, 2017).

### 3.5 Results

There was not a significant difference between gamers and non-gamers in VRNQ scores (see Table 3). The final version of VR-EAL exceeded the parsimonious cut-off for the VRNQ total score, while the alpha and beta versions of VR-EAL did not (see Table 4). Notably, the VRNQ sub-scores of the final version of VR-EAL also exceeded the parsimonious VRNQ cut-offs (see Table 4), while the average duration of the VR sessions (i.e., duration of being immersed) was 62.2 minutes (SD = 5.59) across the 25 participants. The beta version of VR-EAL approached the cut-offs for user experience and game mechanics; however, it was substantially below the cut-offs for in-game assistance and VRISE. The alpha version of VR-EAL was significantly below the cut-offs for every sub-score of VRNQ.

Table 3. Comparison of VRNQ Scores Between Gamers and Non-Gamers

| **VRNQ Scores** | **p-value** | **$BF_{10}$** | **error %** |
|---|---|---|---|
| Total VRNQ | p = 0.631 | 0.402 | 1.052e -4 |
| User Experience | p = 0.289 | 0.546 | 0.001 |
| Game Mechanics | p = 0.459 | 0.429 | 2.003e -4 |
| In-Game Assistance | p = 0.841 | 0.374 | 0.030 |
| VRISE | p = 0.983 | 0.368 | 0.030 |

*\* $BF_{10} > 10$; \*\* $BF_{10} > 30$; \*\*\* $BF_{10} > 100$; No significant differences observed*

Table 4. VRNQ scores for Alpha, Beta, and Final versions of the VR-EAL

| | N | Median (MAD) | Cut-off | Maximum Score |
|---|---|---|---|---|
| Total VRNQ - Alpha Version | 12 | 100 (6) | ≥ 120 | 140 |
| User Experience - Alpha Version | 12 | 25 (2) | ≥ 30 | 35 |
| Game Mechanics - Alpha Version | 12 | 23.5 (3.5) | ≥ 30 | 35 |
| In-Game Assistance - Alpha Version | 12 | 24 (3) | ≥ 30 | 35 |
| VRISE - Alpha Version | 12 | 25.5 (1.5) | ≥ 30 | 35 |
| Total VRNQ - Beta Version | 12 | 109.5 (2.5) | ≥ 120 | 140 |
| User Experience - Beta Version | 12 | 28 (1) | ≥ 30 | 35 |
| Game Mechanics - Beta Version | 12 | 29 (1) | ≥ 30 | 35 |
| In-Game Assistance - Beta Version | 12 | 26 (1) | ≥ 30 | 35 |
| VRISE - Beta Version | 12 | 26 (1) | ≥ 30 | 35 |
| Total VRNQ - Final Version – All | 25 | 128 (5) | ≥ 120 | 140 |
| User Experience - Final Version – All | 25 | 31 (2) | ≥ 30 | 35 |
| Game Mechanics - Final Version – All | 25 | 32 (2) | ≥ 30 | 35 |
| In-Game Assistance - Final Version – All | 25 | 32 (3) | ≥ 30 | 35 |
| VRISE - Final Version – All | 25 | 33 (1) | ≥ 30 | 35 |
| Total VRNQ - Final Version – Gamers | 12 | 129.5 (5) | ≥ 120 | 140 |
| User Experience - Final Version – Gamers | 12 | 32.5 (1.5) | ≥ 30 | 35 |
| Game Mechanics - Final Version – Gamers | 12 | 32 (1.5) | ≥ 30 | 35 |
| In-Game Assistance - Final Version – Gamers | 12 | 32.5 (2) | ≥ 30 | 35 |
| VRISE - Final Version – Gamers | 12 | 33 (1) | ≥ 30 | 35 |
| Total VRNQ - Final Version - Non-Gamers | 13 | 128 (4) | ≥ 120 | 140 |
| User Experience - Final Version - Non-Gamers | 13 | 31 (1) | ≥ 30 | 35 |
| Game Mechanics - Final Version - Non-Gamers | 13 | 31 (2) | ≥ 30 | 35 |
| In-Game Assistance - Final Version - Non-Gamers | 13 | 32 (3) | ≥ 30 | 35 |
| VRISE - Final Version - Non-Gamers | 13 | 33 (2) | ≥ 30 | 35 |

*MAD = Median Absolute Deviation*

According to the adopted nomenclature (i.e., $BF_{10} \leq 1$ indicating no evidence in favor of H1, $BF_{10} > 3$ indicating moderate evidence in favor of H1, $BF_{10} \geq 10$ for H1, and $BF_{10} \geq 100$ indicating extreme evidence for H1) by Marsman & Wagenmakers, (2017), Wagenmakers *et al.*, (2018a), Wagenmakers *et al.*, (2018b), the Bayesian t-test analysis (N = 12) demonstrated significant differences in the VRNQ scores between the final, beta, and alpha versions of the VR-EAL (see Table 5). We observed that the probability of the alternative hypothesis that the VRNQ total score for the final version is greater than the VRNQ total score for the alpha version is 57794 times greater (i.e., $BF_{10} = 57974$; see Table 5) than the probability of H0 (i.e., not being greater). Similarly, the probability of the alternative hypothesis that the VRNQ total score for the final version is greater than the VRNQ total score for the beta version is 855 times greater (i.e., $BF_{10} = 855$; see Table 5) than the probability of H0. Lastly, the probability of the alternative hypothesis that the VRNQ total score for the beta version is greater than the VRNQ total score for the alpha version is 101 times greater (i.e., $BF_{10}$ = 101; see Table 5) than the probability of H0. The remaining alternative hypotheses for the comparisons between the versions of VR-EAL and their probabilities against the corresponding null hypotheses are displayed in Table 5.

Moreover, the final version was substantially better than the alpha version in terms of every sub-score and total score of the VRNQ. The beta version was better than the alpha version in terms of the VRNQ total score as well as the user experience and game mechanics sub-scores. However, there was not a significant difference between the VRNQ in terms of the VRISE or in-game assistance sub-scores. Moreover, the final version was also significantly improved compared to the beta version in terms of the VRNQ total score and all sub-scores. Though, the difference between them was smaller in the game mechanics and user experience sub-scores (see Table 5). Importantly, in the final version of the VR-EAL, all users (N = 25) experienced mild (i.e., 5 in VRNQ) to no VRISE (i.e., 7 in VRNQ), while the vast majority (N=22) experienced very mild (i.e., 6 in VRNQ) to no VRISE (see Figure 8).

Table 5. Bayesian Paired Sample T-Test Results

| Alternative Hypothesis (H1) | | | p-value | $BF_{10}$ | error % |
|---|---|---|---|---|---|
| Total VRNQ - Alpha | < | Total VRNQ - Beta | p < .001 | 101.651*** | ~2.226e -5 |
| Total VRNQ-Alpha | < | Total VRNQ-Final | p < .001 | 57974.267*** | ~9.361e-35 |
| Total VRNQ-Beta | < | Total VRNQ-Final | p < .001 | 855.603*** | ~1.506e-17 |
| User Experience-Alpha | < | User Experience-Beta | p < .001 | 21.221* | ~9.875e -5 |
| User Experience-Alpha | < | User Experience-Final | p < .001 | 681.518*** | ~8.429e-24 |
| User Experience-Beta | < | User Experience-Final | p < .001 | 17.597* | ~2.172e -4 |
| Game Mechanics-Alpha | < | Game Mechanics-Beta | p < .001 | 47.214** | ~1.820e -4 |
| Game Mechanics-Alpha | < | Game Mechanics-Final | p < .001 | 487.798*** | ~2.337e-19 |
| Game Mechanics-Beta | < | Game Mechanics-Final | p < .001 | 17.262* | ~2.288e -4 |
| In-Game Assistance-Alpha | < | In-Game Assistance-Beta | p = .098 | 1.095 | ~9.459e -4 |
| In-Game Assistance-Alpha | < | In-Game Assistance-Final | p < .001 | 224.329*** | ~1.110e-18 |
| In-Game Assistance-Beta | < | In-Game Assistance-Final | p < .001 | 139.994*** | ~5.188e -5 |
| VRISE-Alpha | < | VRISE-Beta | p = .111 | 0.988 | ~0.001 |
| VRISE-Alpha | < | VRISE-Final | p < .001 | 1912.328*** | ~3.643e-24 |
| VRISE-Beta | < | VRISE-Final | p < .001 | 1277.335*** | ~7.819e-21 |

*$BF_{10}$ > 10; ** $BF_{10}$ >30; *** $BF_{10}$ > 100; Alpha = Alpha version of VR-EAL; Beta = Beta version of VR-EAL; Final = Final Version of VR-EAL*

Figure 8. VRISE in the Final Version of VR-EAL

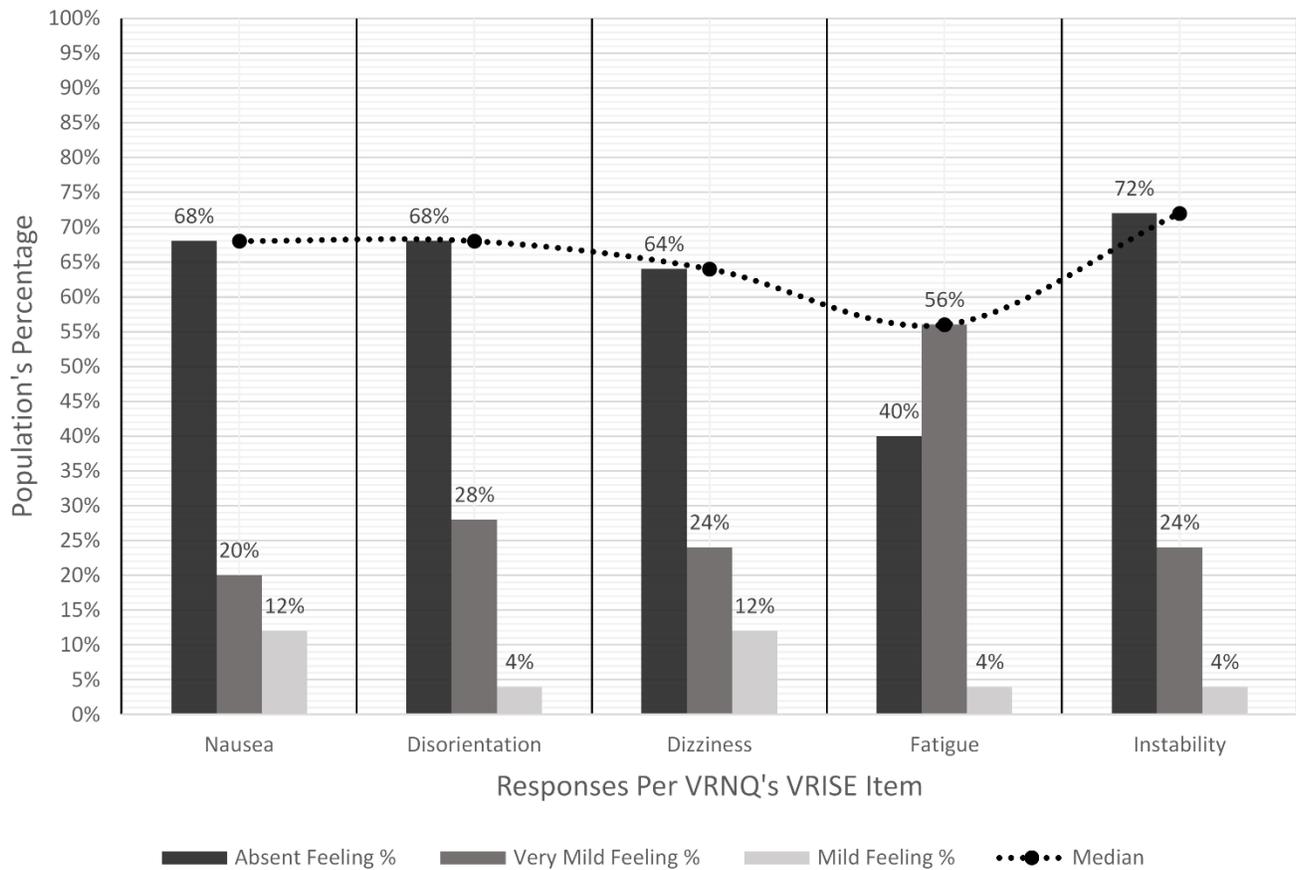

## 4    Discussion

### 4.1    The VR-EAL versions

The present study attempted to develop a cost-effective VR research/clinical software (i.e., VR-EAL) of a high enough quality for implementation in cognitive studies and that does not induce VRISE. The development included three versions of VR-EAL (i.e., alpha, beta and final) until the attainment of these desired outcomes. The alpha version of VR-EAL revealed several limitations. It had low frames per second (fps), which increased the frequency and the intensity of VRISE. Also, the alpha version did not include haptics during the interactions, and the in-game assistance props were low in number. Lastly, the shaders of the 3D models were not converted to VR shaders (i.e., the function of the Lab renderer) and numerous game objects were defined as non-static. As a result, the quality of the graphics was below average, and the fps were substantially below 90 (i.e., 70 -80) which is the lowest threshold for VR software targeting high-end HMDs such as HTC Vive and Oculus Rift. However, the feedback also confirmed that several game mechanics and approaches (e.g., tutorials) were in the right direction, which was encouraging for further VR-EAL development.

The principal improvements in the beta version of VR-EAL were pertinent to the alpha version's shortcomings. The shaders for all the game objects were converted to VR shaders, and several game objects, with which the user does not interact, were defined as static. The fps for the beta version were above 90, though, there were various points where the fps dropped for a couple of seconds. Although these fps drops were brief, their existence negatively affected the users who reported moderate to intense VRISE. Nonetheless, the beta version provided haptic and visual (i.e., highlighters) feedback to the users during the interactions, which further improved the quality of the game mechanics. In addition, the number of in-game aids was dramatically increased (e.g., more signs, labels, and directional arrows) and the duration of the tutorials was substantially prolonged (i.e., the inclusion of more explicit descriptions), which improved the quality of the users' experience. However, while the beta version was an improvement, it still failed to meet the parsimonious cut-offs of the VRNQ.

In the final version of VR-EAL, further improvements were conducted. The programming scripts of VR-EAL were re-assessed and correspondingly refined. Various chunks of code were expressed more compactly. For example, part of the code which had several Boolean values and/or float numbers were replaced by events and delegates (i.e., the features of object-oriented programming languages like C# that have substantially lower costs towards the performance of the software). Furthermore, the lightmapping of the 3D environments of scenes was upgraded by calculating high-resolution lightmaps instead of the medium resolution used in previous versions of VR-EAL. Redundant shadows were also deactivated to improve the performance of VR-EAL without degrading the quality of the graphics.

Moreover, major parts of the 3D environments were baked together (i.e., merged) through the implementation of MeshBaker's predominant functions to significantly reduce the draw-calls of VR-EAL. Interestingly, the result was a stable number of fps during gameplay. Specifically, the final version of VR-EAL has 120–140 fps during gameplay. Lastly, there was an improvement and enrichment of in-game assistance. In the tutorials, video screens and videos were added, which show the user how to use the controllers and perform each task. This visual and procedural demonstration allowed users to learn the respective controls and task trials faster and more effectively. This audio-visual demonstration using videos is feasible in VR since it can integrate the benefits of all mediums (e.g., video, audio, audio-visual). Furthermore, in the storyline scenes, where the user performs the actual tasks, several visual aids were added to provide additional guidance and alleviate confusion (see Figure 3 and 6).

Our results demonstrated that the VRNQ total and sub-scores exceeded the parsimonious cut-offs of the VRNQ for the final VR-EAL version. The improvements pertinent to graphics substantially increased the quality of the user experience, while they almost eradicated VRISE (see Figure 8). This substantial decrease of VRISE also highlights the importance of fps in VR. A developer should use the Unity profiler to check whether the VR software has a steady number of fps during gameplay, which the HMD requires. Also, the final version of VR-EAL appeared to have better in-game assistance and game mechanics. However, there was not any upgrade pertinent to the game mechanics. The increase in the evaluation of the game mechanics probably resulted due to the addition and improvement of in-game aids in both tutorial and storyline scenes. This finding also supports that in-game assistance has a paramount role in VR software. This is especially the case when the software is developed for clinical or research purposes, where the users could be either gamers or non-gamers. The quality of the tutorials and in-game aids should be cautiously designed to ensure the usability of the VR research software. To sum up, the final version of VR-EAL seems to deliver a pleasant testing experience and without the presence of significant VRISE.

## 4.2   VR software development in cognitive sciences

The current study demonstrated the procedure for the development of immersive VR research/clinical software (i.e., VR-EAL) with strong placement and plausibility illusions, which are necessary for collecting genuine responses (i.e., ecological valid) from users (Slater, 2009; Slater *et al.*, 2010; Maister *et al.*, 2015; Pan & Hamilton, 2018). The implementation of good quality 3D models (e.g., objects, buildings, and artificial humans) in conjunction with optimization tools (e.g., Lab Renderer and MeshBaker) facilitated an analogous placement illusion. Also, VR-EAL incorporates naturalistic and ergonomic interactions with the VE facilitated by the VR hardware (e.g., HTC Vive and 6DoF controllers), SDKs (e.g., SteamVR and VRTK), and Unity assets pertinent to spatialized audio (e.g., Steam Audio) and artificial characters' animations (e.g., Salsa3D). These naturalistic and ergonomic interactions with the VE are capable of inducing a robust plausibility illusion.

Furthermore, a predominant concern for the implementation of VR in cognitive sciences is the presence of VRISE (Bohil *et al.*, 2011; de Franca & Soares, 2017; Palmisano *et al.*, 2017), which may compromise health and safety standards (Parsons *et al.*, 2018), as well as the reliability of cognitive (Nalivaiko *et al.*, 2015), physiological (Nalivaiko *et al.*, 2015), and neuroimaging data (Arafat *et al.*, 2018; Gavgani et al, 2018). Equally, the high cost of VR software development may additionally deter the adoption of VR as a research tool in cognitive sciences (Slater, 2018). However, the development of VR-EAL provides evidence that the obstacles above can be surpassed to implement VR software in cognitive sciences effectively.

The users of the final version of VR-EAL reported mild to no VRISE, with the average value in the VRISE sub-score being very mild to no VRISE. Importantly, these reports were offered by the users after spending around 60 minutes uninterrupted in VR. Typically, VRISE are intensified in longer VR sessions (Sharples *et al.*, 2008). However, the utilization of the parsimonious cut-offs from the VRNQ guaranteed the significant alleviation of VRISE, which was also supported by the users' reports. Notably, the results of this study are in line with our previous work (Kourtesis *et al.*, 2019b), where the gaming experience (i.e., gamer or non-gamer) did not affect the responses on the VRNQ. Also, the results support that the gaming experience does not affect the presence or intensity of VRISE in software of adequate quality. Therefore, VR software with technical features similar to VR-EAL would be suitable for implementation in cognitive sciences.

Cognitive scientists already implement computational approaches to investigate cognitive functions at the neuronal and cellular level (Sejnowski, 1988; Farrell & Lewandowsky, 2010; Kriegeskorte & Douglas, 2018), develop computerized neuropsychological tasks compatible with neuroimaging techniques (Mathôt *et al.*, 2011; Peirce, 2009; Peirce et al, 2011), as well as conducting flexible statistical analyses and creating high-quality graphics and simulations (Culpepper & Aguinis, 2011; Revelle, 2011; Stevens, 2017). The development of VR-EAL was achieved by using C# and Unity packages (i.e., SteamVR SDK, VRTK, Lab renderer, MeshBaker, Salsa3D, RandomEyes3D, 3D models, 3D environments, and 3D characters) on the Unity game engine, which is a user-friendly IDE equivalent to OpenSesame, PsychoPy, and MATLAB.

The majority of these Unity packages are cost-free, while the remainder are relatively low-cost, and could be used in future VR software development. Also, the acquisition of VR development skills by cognitive scientists with a background in either psychology or computers science can be realized in a moderately short period. Although, collaboration with a psychologist who has the required knowledge and clinical experience is crucial for a computer scientist with VR skills. Likewise, psychologists should either collaborate with a computer scientist with VR expertise or acquire VR development skills themselves. For the acquisition of VR skills by a computer scientist or a psychologist, there are online and on-campus interdisciplinary modules (e.g., Unity tutorials and documentation, game development courses, programming workshops, and specializations in VR) which further support the feasibility of acquiring the necessary skills. However, training cognitive scientists in VR software development should be prioritized for institutions which aspire to implement VR technologies in their studies. To summarize, this study demonstrated that the development of usable VR research software by a cognitive scientist is viable.

### 4.3 Limitations and future studies

This study, however, has some limitations. The implementation of novel technologies may result in more positive responses towards them (Wells *et al.*, 2010). A future replication of the current results would elucidate this issue. Also, the study did not provide validation of VR-EAL as a neuropsychological tool. Future work will consider validating the VR-EAL against traditional paper-and-pencil and computerized tests of prospective memory, executive function, episodic memory, and attentional processes. A future validation study should also include a larger and more diverse population than the sample in this study. Regarding the quality of VR-EAL, it is not able to induce a strong embodiment illusion. The future version of the VR-EAL should include a VR avatar that corresponds to the user's movements and actions. Also, the integration of better 3D models, environments, and characters may be beneficial, which will additionally improve the quality of placement illusion and the user's experience. Finally, since VR-EAL is ultimately intended for implementation in cognitive neuroscience and neuropsychology, the future version of VR-EAL should include compatibility with eye-tracking measurements and neuroimaging techniques (e.g., event-related potentials measured by electroencephalography).

### 4.4 Conclusion

This study provided guidelines for the development of immersive VR research software that can be implemented in cognitive sciences to improve the ecological validity of the cognitive tasks and automate the administration and scoring of the neuropsychological assessment. The results substantially support the feasibility of the development of low-cost and effective immersive VR software without the presence of VRISE during a 60 minutes VR session by cognitive scientists who have skills in VR software development. Technologically competent cognitive scientists are able to develop cost-effective immersive VR research software that guarantees the safety of the users and the reliability of the collected data (i.e., neuropsychological, physiological, and neuroimaging data).

### 5 Conflict of Interest

The authors declare that the research was conducted in the absence of any commercial or financial relationships that could be construed as a potential conflict of interest.

### 6 Author Contributions

The primary author is the developer of VR-EAL. VR-EAL can be used by a third party by contacting the corresponding author. The primary author had the initial idea and contributed to every aspect of this study. The rest of the authors contributed to the methodological aspects and the discussion of the results.

Table 1. VR-EAL Scenario

| Order | Type | Description |
|-------|------|-------------|
| Scene 1 | *Tutorial* | Basic interactions and navigation |
| Scene 2 | *Tutorial* | Interactive boards (recognition and planning) |
| Scene 3 | Storyline | List of prospective memory tasks, shopping list (immediate recognition), and itinerary (planning) |
| Scene 4 | *Tutorial* | List of mechanics for the prospective memory tasks, prompts, and notes |
| Scene 5 | *Tutorial* | Cooking |
| Scene 6 | Storyline | Prepare breakfast (multi-tasking) and take medication (prospective memory, event-based, short delay) |
| Scene 7 | *Tutorial* | Tutorial: collect items |
| Scene 8 | Storyline | Collect items from the living-room (selective visuospatial attention) and take a chocolate pie out of the oven (prospective memory, event-based, short delay) |
| Scene 9 | *Tutorial* | Interaction with 3D non-player characters |
| Scene 10 | Storyline | Call Rose (prospective memory task, time-based, short delay) |
| Scene 11 | *Tutorial* | Gaze interaction |
| Scene 12 | Storyline | Detect posters on both sides of the road (selective visual attention) |
| Scene 13 | *Tutorial* | Shopping, how to collect the items from the supermarket |
| Scene 14 | Storyline | Collect the shopping list items from the supermarket (delayed recognition) |
| Scene 15 | Storyline | Go to the bakery to collect the carrot cake (prospective memory task, time-based, medium delay) |
| Scene 16 | Storyline | False prompt before going to the library (prospective memory task, event-based, medium delay) |
| Scene 17 | Storyline | Return the red book to the library (prospective memory task, event-based, medium delay) |
| Scene 18 | *Tutorial* | Auditory interaction |
| Scene 19 | Storyline | Detect sounds from both sides of the road (selective auditory attention) |
| Scene 20 | Storyline | False prompt before going back home (prospective memory task, time-based, long delay) |
| Scene 21 | Storyline | When you return home, give the extra pair of keys to Alex (prospective memory task, event-based, long delay) |
| Scene 22 | Storyline | Put away the shopping items and take the medication (prospective memory task, time-based, long delay) |

Table 2. VRNQ minimum and parsimonious cut-offs

| Score | Minimum Cut-offs | Parsimonious Cut-offs |
|---|---|---|
| **User Experience** | $\geq 25/35$ | $\geq 30/35$ |
| **Game Mechanics** | $\geq 25/35$ | $\geq 30/35$ |
| **In-Game Assistance** | $\geq 25/35$ | $\geq 30/35$ |
| **VRISE** | $\geq 25/35$ | $\geq 30/35$ |
| **VRNQ Total Score** | $\geq 100/140$ | $\geq 120/140$ |

*The median of each sub-score and total score should meet the suggested cut-offs to determine that the evaluated VR software is of adequate quality without any significant VRISE. The utilization of the parsimonious cut-offs more robustly supports the suitability of the VR software. Derived from Kourtesis et al., 2019b*

Table 3. Comparison of VRNQ Scores Between Gamers and Non-Gamers

| VRNQ Scores | p-value | $BF_{10}$ | error % |
|---|---|---|---|
| Total VRNQ | $p = 0.631$ | 0.402 | 1.052e -4 |
| User Experience | $p = 0.289$ | 0.546 | 0.001 |
| Game Mechanics | $p = 0.459$ | 0.429 | 2.003e -4 |
| In-Game Assistance | $p = 0.841$ | 0.374 | 0.030 |
| VRISE | $p = 0.983$ | 0.368 | 0.030 |

*\* $BF_{10} > 10$; \*\* $BF_{10} > 30$; \*\*\* $BF_{10} > 100$; No significant differences observed*

Table 4. VRNQ scores for Alpha, Beta, and Final version of VR-EAL

| | N | Median (MAD) | Cut-off | Maximum Score |
|---|---|---|---|---|
| Total VRNQ - Alpha Version | 12 | 100 (6) | ≥ 120 | 140 |
| User Experience - Alpha Version | 12 | 25 (2) | ≥ 30 | 35 |
| Game Mechanics - Alpha Version | 12 | 23.5 (3.5) | ≥ 30 | 35 |
| In-Game Assistance - Alpha Version | 12 | 24 (3) | ≥ 30 | 35 |
| VRISE - Alpha Version | 12 | 25.5 (1.5) | ≥ 30 | 35 |
| Total VRNQ - Beta Version | 12 | 109.5 (2.5) | ≥ 120 | 140 |
| User Experience - Beta Version | 12 | 28 (1) | ≥ 30 | 35 |
| Game Mechanics - Beta Version | 12 | 29 (1) | ≥ 30 | 35 |
| In-Game Assistance - Beta Version | 12 | 26 (1) | ≥ 30 | 35 |
| VRISE - Beta Version | 12 | 26 (1) | ≥ 30 | 35 |
| Total VRNQ - Final Version - All | 25 | 128 (5) | ≥ 120 | 140 |
| User Experience - Final Version - All | 25 | 31 (2) | ≥ 30 | 35 |
| Game Mechanics - Final Version - All | 25 | 32 (2) | ≥ 30 | 35 |
| In-Game Assistance - Final Version - All | 25 | 32 (3) | ≥ 30 | 35 |
| VRISE - Final Version - All | 25 | 33 (1) | ≥ 30 | 35 |
| Total VRNQ - Final Version - Gamers | 12 | 129.5 (5) | ≥ 120 | 140 |
| User Experience - Final Version - Gamers | 12 | 32.5 (1.5) | ≥ 30 | 35 |
| Game Mechanics - Final Version - Gamers | 12 | 32 (1.5) | ≥ 30 | 35 |
| In-Game Assistance - Final Version - Gamers | 12 | 32.5 (2) | ≥ 30 | 35 |
| VRISE - Final Version - Gamers | 12 | 33 (1) | ≥ 30 | 35 |
| Total VRNQ - Final Version - Non-Gamers | 13 | 128 (4) | ≥ 120 | 140 |
| User Experience - Final Version - Non-Gamers | 13 | 31 (1) | ≥ 30 | 35 |
| Game Mechanics - Final Version - Non-Gamers | 13 | 31 (2) | ≥ 30 | 35 |
| In-Game Assistance - Final Version - Non-Gamers | 13 | 32 (3) | ≥ 30 | 35 |
| VRISE - Final Version - Non-Gamers | 13 | 33 (2) | ≥ 30 | 35 |

*MAD = Median Absolute Deviation;*

Table 5. Bayesian Paired Sample T-Test Results

| Alternative Hypothesis (H1) | | | p-value | BF$_{10}$ | error % |
|---|---|---|---|---|---|
| Total VRNQ - Alpha | < | Total VRNQ - Beta | p < .001 | 101.651*** | ~2.226e -5 |
| Total VRNQ-Alpha | < | Total VRNQ-Final | p < .001 | 57974.267*** | ~9.361e-35 |
| Total VRNQ-Beta | < | Total VRNQ-Final | p < .001 | 855.603*** | ~1.506e-17 |
| User Experience-Alpha | < | User Experience-Beta | p < .001 | 21.221* | ~9.875e -5 |
| User Experience-Alpha | < | User Experience-Final | p < .001 | 681.518*** | ~8.429e-24 |
| User Experience-Beta | < | User Experience-Final | p < .001 | 17.597* | ~2.172e -4 |
| Game Mechanics-Alpha | < | Game Mechanics-Beta | p < .001 | 47.214** | ~1.820e -4 |
| Game Mechanics-Alpha | < | Game Mechanics-Final | p < .001 | 487.798*** | ~2.337e-19 |
| Game Mechanics-Beta | < | Game Mechanics-Final | p < .001 | 17.262* | ~2.288e -4 |
| In-Game Assistance-Alpha | < | In-Game Assistance-Beta | p = .098 | 1.095 | ~9.459e -4 |
| In-Game Assistance-Alpha | < | In-Game Assistance-Final | p < .001 | 224.329*** | ~1.110e-18 |
| In-Game Assistance-Beta | < | In-Game Assistance-Final | p < .001 | 139.994*** | ~5.188e -5 |
| VRISE-Alpha | < | VRISE-Beta | p = .111 | 0.988 | ~0.001 |
| VRISE-Alpha | < | VRISE-Final | p < .001 | 1912.328*** | ~3.643e-24 |
| VRISE-Beta | < | VRISE-Final | p < .001 | 1277.335*** | ~7.819e-21 |

*$BF_{10} > 10$; ** $BF_{10} > 30$; *** $BF_{10} > 100$; Alpha = Alpha version of VR-EAL; Beta = Beta version of VR-EAL; Final = Final Version of VR-EAL*